\documentclass[reprint, twocolumn, floatfix, superscriptaddress, preprintnumbers]{revtex4-2}

\usepackage{mathtools}
\usepackage{physics}
\usepackage{amsmath}
\usepackage{tabto}
\usepackage{graphicx}
\usepackage{dcolumn}
\usepackage{bm}
\usepackage{comment}
\usepackage{algorithm} 
\usepackage{algpseudocode}
\graphicspath{ {images/} }

\usepackage{hyperref}
\usepackage{cleveref}
\usepackage{float}
\usepackage[caption=false]{subfig}
\usepackage{color}
\usepackage[dvipsnames]{xcolor}

\begin{document}
\title{Robust quantum control with disorder-dressed evolution}

\author{Tenzan Araki}
\email{taraki@ethz.ch}
\affiliation{Quantum Engineering, Department of Information Technology and Electrical Engineering,  ETH Z\"{u}rich, 8092 Z\"{u}rich, Switzerland}
\affiliation{Theoretical Quantum Physics Laboratory, Cluster for Pioneering Research, RIKEN, Wako, Saitama 351-0198, Japan}
\author{Franco Nori}
\affiliation{Theoretical Quantum Physics Laboratory, Cluster for Pioneering Research, RIKEN, Wako, Saitama 351-0198, Japan}
\affiliation{Center for Quantum Computing, RIKEN, Wako, Saitama 351-0198, Japan}
\affiliation{Department of Physics, University of Michigan, Ann Arbor, Michigan 48109-1040, USA}
\author{Clemens Gneiting}
\email{clemens.gneiting@riken.jp}
\affiliation{Theoretical Quantum Physics Laboratory, Cluster for Pioneering Research, RIKEN, Wako, Saitama 351-0198, Japan}
\affiliation{Center for Quantum Computing, RIKEN, Wako, Saitama 351-0198, Japan}

\date{\today}
\begin{abstract}
The theory of optimal quantum control serves to identify time-dependent control Hamiltonians that efficiently produce desired target states. As such, it plays an essential role in the successful design and development of quantum technologies. However, often the delivered control pulses are exceedingly sensitive to small perturbations, which can make it hard if not impossible to reliably deploy these in experiments. Robust quantum control aims at mitigating this issue by finding control pulses that uphold their capacity to reproduce the target states even in the presence of pulse perturbations. However, finding such robust control pulses is generically hard, since the assessment of control pulses requires the inclusion of all possible distorted versions in the evaluation. Here we show that robust control pulses can be identified based on disorder-dressed evolution equations. The latter capture the effect of disorder, which here stands for the pulse perturbations, in terms of quantum master equations describing the evolution of the disorder-averaged density matrix. In this approach to robust control, the purities of the final states indicate the robustness of the underlying control pulses, and robust control pulses are singled out if the final states are pure (and coincide with the target states). We show that this principle can be successfully employed to find robust control pulses. To this end, we adapt Krotov's method for disorder-dressed evolution, and demonstrate its application with several single-qubit control tasks.
\end{abstract}

\preprint{\textsf{published in Phys.~Rev.~A~{\bf 107}, 032609 (2023)}}

\maketitle

\section{\label{sec:1}Introduction}

The increasingly precise control of individual quantum systems has brought into reach the active harnessing of quantum properties towards quantum technologies with a tangible quantum advantage. Potential applications range from quantum sensing \cite{degen2017}, to quantum communication \cite{chen1865, muralidharan2016}, quantum simulation \cite{buluta2009, georgescu2014, monroe2021}, and quantum computation \cite{divincenzo2000, nielsen2010}. Promising platforms \cite{buluta2011} that are currently under intense development include, for instance, superconducting circuits, trapped ions, quantum dots, ultracold atoms in optical lattices, and nitrogen vacancy centers.

Besides shielding devices from the detrimental effect of environmental decoherence, the accurate and efficient control of systems' quantum dynamics is an indispensable prerequisite for the successful deployment of quantum technologies. This is the objective of optimal quantum control, which aims at identifying optimal control pulses such that the resulting Hamiltonians generate a desired quantum evolution \cite{Dong2010quantum, brif2010, glaser2015, koch2016, dalessandro2021}. Such control pulses, which often correspond to pulses of external electromagnetic fields applied to the quantum systems, lie, for instance, at the heart of the realization of quantum logic gates in the circuit model of quantum computation.

While optimal control pulses can, in rare cases, be determined analytically, one typically must resort to numerical means \cite{pontryagin1987, glaser2015}. Numerical approaches include, e.g., the Krotov \cite{sklarz2002, palao2003, reich2012, morzhin2019}, the GRAPE (GRadient Ascent Pulse Engineering) \cite{khaneja2005}, and the CRAB (Chopped Random-Basis) \cite{doria2011, caneva2011} algorithms. Various experiments have successfully deployed optimal control pulses obtained through these methods \cite{lovecchio2016, vanfrank2016, heeres2017, heck2018, feng2018}. However, such numerically obtained control pulses in general prohibit a transparent interpretation, which makes it hard if not impossible to assess their performance under perturbations.

Under realistic experimental conditions, we must expect that imprecise device control and uncontrolled external influences, e.g., stray fields, limit the accurate implementation of control pulses, resulting in deviations from the desired dynamics. {\it Robust quantum control} aims to mitigate the impact of such noise and disorder by identifying control pulses that uphold their performance even under the presence of perturbations (see, e.g.,~\cite{zhang1994, Li2006control, Montangero2007robust, Leghtas2011adiabatic, Ruths2011multidimensional, ruschhaupt2012, chen2013, Daems2013robust, goerz2014b, Chen2014sampling, Dong2015robust, dong2016, VanDamme2017robust, sakai2019, ball2021, carvalho2021, boxili2022}). Robust control thus relies on the insight that control pulses are not unique, which gives us the freedom to further select them for robustness.

Various approaches to robust quantum control have been proposed, including those adapted from classical control theory \cite{dhelon2006, james2008, dong2009}. A common and intuitive strategy to numerically find robust control pulses relies on sampling-based ``ensemble optimization'', where the average fidelities over randomly drawn ensembles of perturbed pulses are compared for different unperturbed pulses \cite{Kobzar2004exploring, goerz2014b, Chen2014sampling, Dong2015robust}; robust pulses are then identified as those which maximize the average fidelity with the target state. Analytical robust control solutions for special cases have been developed, e.g., in the context of ``shortcuts to adiabaticity'' \cite{ruschhaupt2012, Guery2019shortcuts}.

Here, we propose a deterministic method for the identification of robust control pulses based on the formalism of disorder-dressed quantum evolution. In this framework, which holds in the perturbative limit of weak pulse perturbations, an evolution equation for the disorder-averaged quantum state is formulated, where the disorder here stands for the pulse perturbations \cite{Gneiting2017quantum, gneiting2020} (earlier versions are found in \cite{gneiting2016a, gneiting2016b}, and applications to condensed matter systems are described, e.g., in \cite{Gneiting2017disorder, Gneiting2018lifetime, Gneiting2019disorder, Han2019helical}). Even if each disorder realization follows a coherent quantum evolution (i.e., is described as an isolated quantum system, where a pure state remains pure), the dynamics of the disorder-averaged state is in general incoherent, and hence is captured by an (in general non-Markovian) quantum master equation. The loss of coherence, or equivalently purity, of the disorder-averaged state then reflects the degree of divergence among the different disorder realizations. This directly leads to the key insight for our application to robust control: \textit{A control pulse can be identified as robust, if the purity of the disorder-averaged state revives when the control pulse approaches its completion.} We use this principle in order to optimize control pulses based directly on the disorder-dressed evolution (instead of the Schr\"odinger equation); pulses that are optimized this way are automatically robust, removing the need for a separate ensemble search for robustness.

To formulate our approach, we first adapt, in \Cref{sec:2}, the disorder-dressed master equation (DDME) \cite{gneiting2020} to the context of optimal control, where the disorder describes small perturbations of the control pulse. This can be seen as a generalization to the DDME derived in \cite{gneiting2020}, where we now also include time-dependent pulse perturbations. In \Cref{sec:3}, we then present an algorithm, based on the well-known Krotov method, which numerically finds optimal control pulses that maximize the final-time fidelity between the disorder-averaged state and the pure target state. As we will show, the standard Krotov method must now be generalized to take the disorder-induced incoherent contributions to the DDME into account. While similar in spirit to ensemble optimization, there is no explicit average over random disorder realizations involved, as the DDME inherently describes the effect of the disorder average. In \Cref{sec:4}, we then demonstrate the viability of our optimization algorithm with three paradigmatic single-qubit operations that are commonly performed as quantum logic gates: a $Z$ gate, an $X$ gate, and a Hadamard gate. In each example, we observe the purity revivals predicted by the DDME-optimized control pulses and we show how this results in significantly increased target-state fidelities compared to control pulses that are naively optimized based on the Schr\"{o}dinger equation (not taking pulse perturbations into consideration).

\section{\label{sec:2}Disorder-dressed Evolution from Pulse Perturbations}

We now derive the disorder-dressed master equation for general time-dependent Hamiltonians and disorder potentials. The latter are subsequently identified with pulse perturbations in the context of optimal control. The (in general) mixed density matrix that solves this equation describes the ensemble average over a collection of pulse perturbations, and thus comprises the disorder effect statistically robustly in a single quantum state. We assume that the effect of pulse perturbations dominates over environmental decoherence, and hence single disorder realizations can be described as closed quantum systems.

We model a disordered quantum system as an ensemble $\{(\hat{H}_\epsilon(t), p_\epsilon)\}$ of perturbed Hamiltonians $\hat{H}_\epsilon(t)$, each associated with its corresponding probability of occurrence $p_\epsilon$, where $\epsilon$ denotes a discrete or continuous index over the set of disorder realizations. For the sake of concreteness, we consider, unless specified otherwise, $\epsilon$ to be continuous.

We derive the DDME following \cite{gneiting2020}, but now generalized to time-dependent Hamiltonians and perturbations of the form
\begin{equation}
    \hat{H}_\epsilon(t) = \hat{\bar{H}}(t)+\hat{V}_\epsilon(t),
    \label{def. H_eps}
\end{equation}
where the mean Hamiltonian $\hat{\bar{H}}(t) \equiv \int\,d\epsilon\,p_\epsilon\hat{H}_\epsilon(t)$ represents the desired Hamiltonian giving rise to the intended dynamics and the deviations $\hat{V}_\epsilon(t)$ represent time-dependent perturbations (usually denoted disorder potentials) satisfying 
\begin{equation}
    \int\,d\epsilon\,p_\epsilon\hat{V}_\epsilon(t) = 0 \hspace{0.5 cm} \forall t.
    \label{eq. balanced deviation}
\end{equation}
We first derive a general form of the DDME based on these definitions, and later specify $\hat{\bar{H}}(t)$ and $\hat{V}_\epsilon(t)$ to arrive at the DDME that can be interpreted in the context of pulse perturbations in optimal control.

A single realization $\hat{\rho}_\epsilon(t)$ within the ensemble follows a closed-system evolution and can thus be described by the von Neumann equation,
\begin{equation}
    \partial_t\hat{\rho}_\epsilon(t) = -\frac{i}{\hbar}[\hat{H}_\epsilon(t), \hat{\rho}_\epsilon(t)]
    \label{eq. rho_eps vN} ,
\end{equation}
and all realizations evolve from the same initial state $\hat{\rho}_\epsilon(0) = \hat{\rho}_0$. To discuss formal solutions of \eqref{eq. rho_eps vN}, we introduce the time evolution operator for some time-dependent Hamiltonian $\hat{H}(t)$,
\begin{equation}
    \hat{U}(t_\text{f},t_\text{i}) = \mathcal{T}\exp{-\frac{i}{\hbar}\int^{t_\text{f}}_{t_\text{i}}\,dt'\,\hat{H}(t')},
    \label{def. time evolution op}
\end{equation}
where $\mathcal{T}$ denotes time ordering and we use the shorthand notation $\hat{U}(t_\text{f}) \equiv \hat{U}(t_\text{f},0)$. With this convention, the time-evolution operators generated by the Hamiltonians $\hat{\bar{H}}$ and $\hat{H}_\epsilon(t)$ are denoted by $\hat{\bar{U}}(t_\text{f},t_\text{i})$ and $\hat{U}_\epsilon(t_\text{f},t_\text{i})$ from here on.

We seek an evolution equation for the disorder-averaged quantum state
\begin{equation}
    \hat{\bar{\rho}}(t) \equiv \int\,d\epsilon\,p_\epsilon\hat{\rho}_\epsilon(t) = \int\,d\epsilon\,p_\epsilon\hat{U}_\epsilon(t)\hat{\rho}_0\hat{U}^\dagger_\epsilon(t),
    \label{def. disorder-averaged state}
\end{equation}
which statistically describes the effect of the perturbations without resorting to individual disorder realizations. To this end, we define the individual offsets of $\hat{\rho}_\epsilon(t)$ from the disorder-averaged state, denoted by $\Delta\hat{\rho}_\epsilon(t)$, so that 
\begin{equation}
    \hat{\rho}_\epsilon(t) = \hat{\bar{\rho}}(t) + \Delta\hat{\rho}_\epsilon(t).
    \label{def. offsets}
\end{equation}
By inserting \eqref{def. H_eps} and \eqref{def. offsets} into \eqref{eq. rho_eps vN}, we obtain
\begin{align}
\begin{split}
    \partial_t\hat{\rho}_\epsilon(t) =& -\frac{i}{\hbar}[\hat{\bar{H}}(t), \hat{\bar{\rho}}(t)] - \frac{i}{\hbar}[\hat{V}_\epsilon(t), \hat{\bar{\rho}}(t)]\\ &-\frac{i}{\hbar}[\hat{\bar{H}}(t), \Delta\hat{\rho}_\epsilon(t)] - \frac{i}{\hbar}[\hat{V}_\epsilon(t), \Delta\hat{\rho}_\epsilon(t)].
    \end{split}
    \label{ddme der. rewrite ito offsets}
\end{align}
Taking the ensemble average as in \eqref{def. disorder-averaged state} then yields
\begin{equation}
    \partial_t \hat{\bar{\rho}}(t) = -\frac{i}{\hbar}[\hat{\bar{H}}(t), \hat{\bar{\rho}}(t)] - \frac{i}{\hbar}\int\,d\epsilon\,p_\epsilon[\hat{V}_\epsilon(t), \Delta\hat{\rho}_\epsilon(t)].
    \label{ddme der. after ensemble average}
\end{equation}
This shows that the dynamics of the disorder-averaged state is coupled to the individual offsets $\Delta\hat{\rho}_\epsilon(t)$ caused by the disorder potentials $\hat{V}_\epsilon(t)$, which gives rise to an incoherent evolution term that can generally lead to a loss of coherence. The evolution equations for the offsets $\Delta\hat{\rho}_\epsilon(t)$ can be obtained by taking the time derivatives in \eqref{def. offsets} and inserting \eqref{ddme der. after ensemble average}, yielding
\begin{align}
    \begin{split}
    &\partial_t\Delta \hat{\rho}_\epsilon(t) +\frac{i}{\hbar}[\hat{H}_\epsilon(t),\Delta\hat{\rho}_\epsilon(t)]\\
    =& -\frac{i}{\hbar}[\hat{V}_\epsilon(t), \hat{\bar{\rho}}(t)] + \frac{i}{\hbar}\int\,d\epsilon'\,p_{\epsilon'}[\hat{V}_{\epsilon'}(t), \Delta\hat{\rho}_{\epsilon'}(t)].
    \label{ddme der. eq for offsets}
    \end{split}
\end{align}

In the short-time limit, the offsets to the disorder-averaged state are sufficiently small so that we can approximate $\Delta \hat{\rho}_\epsilon (t) \approx 0$. By inserting this into \eqref{ddme der. eq for offsets} and integrating, we immediately obtain $\Delta \hat{\rho}_\epsilon(t) = -\frac{i\,t}{\hbar}[\hat{V}_\epsilon(t),\hat{\bar{\rho}}(t)]$, which can then be substituted into \eqref{ddme der. after ensemble average} to recover the short-time master equation derived in \cite{gneiting2016a}, now generalized to the time-dependent case.

The source terms on the right-hand side of \eqref{ddme der. eq for offsets} exhibit contributions from the disorder-averaged state and from the coupling to the offsets of other disorder realizations. With the initial condition $\Delta\hat{\rho}_\epsilon(0) = 0$, the formal solution of \eqref{ddme der. eq for offsets} reads, using Green's formalism,
\begin{align}
    \begin{split}
    \Delta\hat{\rho}_\epsilon(t) =& \int^t_0\,dt'\,\hat{U}_\epsilon(t,t')\bigg{\{}-\frac{i}{\hbar}[\hat{V}_\epsilon(t'),\hat{\bar{\rho}}(t')]\\
    &+\frac{i}{\hbar}\int\,d\epsilon'p_{\epsilon'}[\hat{V}_{\epsilon'}(t'),\Delta\hat{\rho}_{\epsilon'}(t')]\bigg{\}}\hat{U}^\dagger_\epsilon(t,t').
    \label{ddme der. after Green's}
    \end{split}
\end{align}
For the control problem to be meaningful, we can assume that the disorder is weak compared to the intended Hamiltonian, and hence we can approximate \eqref{ddme der. after Green's} to first order in $\hat{V}_\epsilon(t)$, which includes $\hat{U}_\epsilon(t,t') \approx \hat{\bar{U}}(t,t')$, so that
\begin{equation}
    \Delta\hat{\rho}_\epsilon(t) \approx -\frac{i}{\hbar}\int^t_0\,dt'\,[\hat{\Tilde{V}}_\epsilon(t,t'),\hat{\bar{\rho}}(t)],
    \label{ddme der. 1st order offset}
\end{equation}
where we defined $\hat{\Tilde{V}}_\epsilon(t,t') \equiv \hat{\bar{U}}(t,t')\hat{V}_\epsilon(t')\hat{\bar{U}}^\dagger(t,t')$.

Finally, by substituting \eqref{ddme der. 1st order offset} into \eqref{ddme der. after ensemble average}, we obtain the general form of the DDME,
\begin{align}
    \partial_t\hat{\bar{\rho}}(t) =& -\frac{i}{\hbar}[\hat{\bar{H}}(t),\hat{\bar{\rho}}(t)] \label{eq. general ddme} \\
    &-\frac{1}{\hbar^2}\int\,d\epsilon\,p_\epsilon\int^t_0\,dt'\,[\hat{V}_\epsilon(t),[\hat{\Tilde{V}}_\epsilon(t,t'),\hat{\bar{\rho}}(t)]].\nonumber
\end{align}
This equation, which holds for general time-dependent Hamiltonians, will be the basis for our analysis of robust quantum control. Apart from the assumption that the disorder potentials can be treated perturbatively, the derivation is general, in particular with respect to the dimension of the system and the control pulses. In contrast to the disorder-dressed master equation in the static limit (i.e., time-independent Hamiltonians and correlations within individual disorder realizations are temporally unbounded), derived in \cite{gneiting2020}, the evolution \eqref{eq. general ddme} allows for time-dependent intended Hamiltonians and disorder potentials, thus broadening the scope of analysis to time-dependent control pulses and perturbations with possibly finite temporal correlations.

Let us remark that, similar to the time-independent case derived in \cite{gneiting2020}, the evolution equation \eqref{eq. general ddme} can be given the algebraic structure of the Lindblad equation, which then allows one to assess the non-Markovian nature of the evolution and its positivity.

By interpreting the disorder in terms of pulse perturbations, we can now write the intended Hamiltonian and the disorder potentials explicitly in terms of control pulses,
\begin{equation}
    \hat{\bar{H}}(t) = \hat{H}_0 + \sum^M_{m=1} f_m(t)\hat{H}_m ,
    \label{def. Hamiltonian with control pulse}
\end{equation}
and their associated perturbations,
\begin{equation}
    \hat{V}_\epsilon(t) = \sum^M_{m=1} g_{\epsilon,m}(t)\hat{H}_m ,
    \label{def. disorder potential with pulse perturbation}
\end{equation}
where $M$ denotes the number of control pulses. Here, $\hat{H}_0$ represents the drift Hamiltonian and $\{\hat{H}_m\}_{m=1}^M$ is a set of control Hamiltonians with associated control pulses $\{f_m(t)\}_{m=1}^M$. Each of the control pulses is subject to a small time-dependent perturbation $g_{\epsilon,m}(t) \ll f_m(t)$, where both $g_{\epsilon,m}(t)$ and $f_m(t)$ are considered to be real functions in this work. By inserting the resulting disordered Hamiltonian into \eqref{eq. general ddme}, we obtain
\begin{align}
    \partial_t\hat{\bar{\rho}}(t) &= -\frac{i}{\hbar}[\hat{\bar{H}}(t),\hat{\bar{\rho}}(t)] \label{eq. ddme}\\
    &-\frac{1}{\hbar^2}\sum^{M}_{m,n=1}\int^t_0\,dt'\,C_{m,n}(t,t')[\hat{H}_{m},[\hat{\Tilde{H}}_{n}(t,t'),\hat{\bar{\rho}}(t)]],\nonumber
\end{align}
where $\hat{\Tilde{H}}_{n}(t,t') \equiv \hat{\bar{U}}(t,t')\hat{H}_{n}\hat{\bar{U}}^{\dagger}(t,t')$ and $C_{m,n}(t,t')$ represents the correlations between the perturbations of the pulses $m$ and $n$, given by
\begin{equation}
    C_{m,n}(t,t') \equiv \int d\epsilon\,p_\epsilon g_{\epsilon,m}(t)g_{\epsilon,n}(t').
    \label{def. correlation function}
\end{equation}
Note that, while the first term of \eqref{eq. ddme} corresponds to the unitary evolution generated by the intended Hamiltonian, the presence of disorder gives rise to effective decoherence, as described by the second term. In the remainder, we refer to \eqref{eq. ddme} as the DDME.

We remark that the perturbative nature of the DDME implies a finite temporal validity range that depends on the amplitudes of the pulse perturbations. Outside its validity range, the solution of the DDME ceases to be a good approximation to the disorder-averaged quantum state (\ref{def. disorder-averaged state}), and eventually may even become unphysical, i.e., exhibit negative eigenvalues. In the context of optimal control, the prerequisite that the disorder-induced deviations remain small at the target time (an essential condition for high-fidelity applications) guarantees that the DDME operates within its limits of validity. Indeed, in the numerical examples considered below, we find excellent agreement between the solution of the DDME and the respective brute-force ensemble-averaged states.

We can recover the static-limit quantum master equation derived in \cite{gneiting2020}, if we consider both $\hat{\bar{H}}(t)$ and $\hat{V}_\epsilon(t)$ in \eqref{eq. general ddme} to be constant in time. In the context of optimal and robust quantum control, however, the possibility of time-dependent control pulses is imperative. When assuming a time-constant correlation function while keeping $\hat{\bar{H}}(t)$ time dependent, Eq.~\eqref{eq. ddme} allows for the analysis of time-dependent control pulses under static pulse perturbations.

In the opposite limit of vanishing correlation time (for simplicity, we also assume vanishing correlations among the control pulses), $C_{m,n}(t,t') = \alpha\delta(t-t')\delta_{mn}$ $\forall m,n \in \{1,2,...,M\}$, with $\alpha > 0$, the DDME reduces to
\begin{equation}
    \partial_t\hat{\bar{\rho}}(t) = -\frac{i}{\hbar}[\hat{\bar{H}}(t),\hat{\bar{\rho}}(t)]-\frac{\alpha}{2\hbar^2}\sum^{M}_{m=1}[\hat{H}_{m},[\hat{H}_{m},\hat{\bar{\rho}}(t)]],
    \label{eq. Gaussian white noise DDME}
\end{equation}
which agrees with the quantum master equation for Gaussian white noise considered, e.g., in \cite{kiely2021}. It is instructive to convert \eqref{eq. Gaussian white noise DDME} into Lindblad form
\begin{equation}
\partial_t\hat{\bar{\rho}}(t) = -\frac{i}{\hbar}[\hat{\bar{H}}(t),\hat{\bar{\rho}}(t)] +\frac{\alpha}{\hbar^2}\sum^{M}_{m=1}\mathcal{L}\big{(}\hat{H}_{m}\big{)}\,\hat{\bar{\rho}}(t),\label{eq. Gaussian white noise DDME in Lindblad form}
\end{equation}
where $\mathcal{L}\big{(}\hat{L}\big{)}\,\hat{\rho} = \hat{L}\hat{\rho}\hat{L}^{\dagger} - \frac{1}{2}\hat{L}^{\dagger}\hat{L}\hat{\rho} - \frac{1}{2}\hat{\rho}\hat{L}^{\dagger}\hat{L}$. This master equation is manifestly Markovian, in which case the Hermitian Lindblad operators $\hat{H}_m$ can never increase the state purity. This shows that the purity resurgences which characterize robust control pulses cannot be observed in the limit of vanishing temporal correlations, and pulse optimization can at best minimize the purity loss in this limit. Only finite temporal correlation times can give rise to the non-Markovian behavior that empowers robust quantum control.

\section{\label{sec:3}Krotov-based Optimization}

By the definition of $\hat{\bar{\rho}}(t)$, it follows directly that the fidelity with the target state $\hat{\rho}_{\text{targ}}$ is equal to unity at time $t$ if and only if the fidelity between $\hat{\rho}_\epsilon(t)$ and the target state is equal to unity for all $\epsilon$. Therefore, by maximizing the fidelity between a target state and the disorder-averaged state at some specified final time $T$, one can obtain a set of control pulses that drive the initial state to the target state robustly under the influence of disorder.

The purity of a disorder-averaged state, defined by $P(\hat{\bar{\rho}}(t)) = \Tr{\hat{\bar{\rho}}(t)^2}$, is also equal to unity at the final time if its fidelity with a pure target state is equal to unity. In the context of pulse perturbations, this intuitively corresponds to the situation where the closed evolution reaches the target state regardless of any pulse perturbation that may occur in the disorder model. Thus, for a given set of control pulses and a model of disorder, one can use the purity at the final time of the disorder-averaged state driven by these control pulses as a measure of robustness. Similarly to \cite{gneiting2020}, one can convert \eqref{eq. general ddme} into Lindblad form and notice that the presence of negative decoherence rates can give rise to a resurgence of coherence in the system. With a robust set of control pulses, the purity may initially decay at times $t > 0$ due to ensemble averaging, but then increase again before $t = T$ so that it reaches unity at the final time. We stress that, under the strictly unital dynamics described by the disorder average, such purity increases are necessarily an indication of the non-Markovian nature of the evolution.

Here we develop a pulse-optimization algorithm that maximizes the fidelity between a pure target state and the disorder-averaged state, $F(\hat{\rho}_{\text{targ}}, \hat{\bar{\rho}}(t)) = \Tr{\sqrt{\hat{\rho}_{\text{targ}}^{1/2}\hat{\bar{\rho}}(t)\hat{\rho}_{\text{targ}}^{1/2}}}$, evaluated at the final time of the disorder-dressed evolution. Starting from a set of control pulses that drive an initial state to the target state with fidelity equal to unity in the disorderless limit, we iteratively optimize the pulse shapes over each of their discretized time steps as we reintroduce disorder. The algorithm is inspired by the linear variant of Krotov's method, which is a standard optimal quantum control algorithm that is usually applied to closed quantum systems following linear evolution equations \cite{morzhin2019}. However, Krotov's method has also been generalized to nonunitary evolutions by considering the density operator as a vector in Liouville space and replacing the Hamiltonian by a Liouvillian \cite{bartana1997, schmidt2011, goerz2014a, basilewitsch2019}. Similarly, the algorithm described here generalizes to disorder-dressed evolutions by replacing the usual von Neumann equation with the DDME.

Krotov's method is an iterative optimization algorithm, for which the pulse update rule is designed to achieve, by construction, monotonic convergence of its cost functional. We consider here the linear variant of the algorithm, where the guarantee for monotonic convergence may be lost in some control problems, but which often still converges for an appropriate choice of step size.

To specify the quantum evolution to be solved with the DDME in each iteration, the algorithm requires the input of an initial state $\hat{\rho}_0$, a set of initial guess pulses $\{f_m^{\text{guess}}(t)\}_{m=1}^M$, drift and control Hamiltonians, and temporal correlation functions governing the disorder or noise suffered by the control pulses~[cf.~\eqref{def. correlation function}]. The guess pulses will only be used in the first iteration, after which the control pulses will be repeatedly updated. In order to harness the disorder-averaged state as the solution of the DDME to obtain the updated control pulses, the algorithm further requires a target state $\hat{\rho}_\text{targ}$, a set of inverse Krotov step sizes $\{\lambda_m\}_{m=1}^M$, and a set of update shape functions $\{S_m(t)\}_{m=1}^M$ that can be used to ensure boundary conditions on the control pulses, where $S_m(t) \in [0,1]$ $\forall m$. The cost functional is given by \cite{bartana1997, basilewitsch2019}
\begin{subequations}
\begin{equation}
    \begin{split}
    J(\{f_m^{(i)}(t)\}_{m=1}^M) =& \,J_\text{T}(\{f_m^{(i)}(t)\}_{m=1}^M)\\
    &+ \sum^M_{m=1} \lambda_m \int^T_0dt\,\frac{\big{[}\Delta f^{(i)}_m(t)\big{]}^2}{S_m(t)},
    \end{split}
    \label{def. cost functional}
\end{equation}
where
\begin{equation}
    J_\text{T}(\{f_m^{(i)}(t)\}_{m=1}^M) = 1 - \Tr{\hat{\rho}_{\text{targ}}\hat{\bar{\rho}}^{(i)}(T)}
    \label{def. J_T}
\end{equation}
and
\begin{equation}
    \Delta f^{(i)}_m(t) \equiv f^{(i)}_m(t) - f_m^{\text{ref}}(t)
    \label{def. cost for pulse}
\end{equation}
\end{subequations}
for some reference pulse $f_m^{\text{ref}}(t)$ to the $m$th control pulse and we use superscripts to denote the iteration number $i \in \{0,1,2,...\}$ with $f_m^{(0)} (t) \equiv f_m^{\text{guess}}(t)$ $\forall m$. In this work we use the standard choice $f_m^{\text{ref}}(t) = f^{(i-1)}_m(t)$ \cite{eitan2011}. $J_\text{T}$ corresponds to the infidelity and is the main part of $J$ that we would like to minimize; the second term of $J$ is a running cost on the control pulses, which is necessary for the derivation of the Krotov update step.

Let us express the right-hand side of the DDME as a superoperator $\mathcal{K}$ that depends on the upper limit $t$ of the time integral and all control pulses $\{f_m(t')\}_{m=1}^M$ $\forall \, 0 \leq t' \leq t$, acting on $\hat{\bar{\rho}}(t)$ so that $\partial_t \hat{\bar{\rho}}(t) = \mathcal{K}(t, \{f_m(t')\}_{m=1}^M)\,\hat{\bar{\rho}}(t)$. The algorithm then involves solving the costate $\hat{\bar{\chi}}(t)$ from the final value problem
\begin{align}
\begin{cases}
    \,\partial_t\hat{\bar{\chi}}^{(i)}(t) = - \mathcal{K}^\dagger(t, \{f^{(i)}_m(t')\}_{m=1}^M)\,\hat{\bar{\chi}}^{(i)}(t)\\
    \,\hat{\bar{\chi}}^{(i)}(T) = \hat{\rho}_{\text{targ}}.
\end{cases}
    \label{eq. co-state FVP}
\end{align}
Note that the time integral in the DDME is still evaluated from $0$ to $t$, even though the equation is solved backward. Within the algorithm, this corresponds to first solving for
\begin{equation}
    \hat{\eta}_{m,n}(t) \equiv \int^t_0\,dt'\,C_{m,n}(t,t')\hat{\Tilde{H}}_{n}(t,t'),
    \label{def. eta}
\end{equation}
cf.~\eqref{eq. ddme}, and then solving \eqref{eq. co-state FVP} backward by treating it as a time-local equation that depends on $\hat{\eta}_{m,n}(t)$.

In practice, the disorder-averaged state is evaluated on a discretized time grid, where $t_s = s\Delta t$ for $s \in \{0,1,...,N_{\text{T}}\}$ with uniform spacing $\Delta t \equiv \frac{T}{N_{\text{T}}}$. Every control pulse is then evaluated on an interleaved time grid such that $f_{m,(k)} \equiv f_m(\tilde{t}_{k-1})$ for $k \in \{1,2,...,N_{\text{T}}\}$ and $\tilde{t}_{k-1} \equiv \frac{t_{k-1}+t_k}{2}$. To avoid confusion, we use subscripts with square brackets to denote evaluation on the former time grid and round brackets for the latter. We introduce, based on first-order Lie-Trotter decomposition, a superoperator
\begin{equation}
    \mathcal{V}_{(k)} \approx \exp{\Delta t \, \mathcal{K}_{(k),\{f_{m,(k')}\}}} \hspace{0.5 cm} \forall \,1 \leq k' \leq k
    \label{def. DDME superoperator}
\end{equation}
such that $\hat{\bar{\rho}}_{[k]} = \mathcal{V}_{(k)}\,\hat{\bar{\rho}}_{[k-1]}$, where $\mathcal{K}_{(k),\{f_{m,(k')}\}} \equiv \mathcal{K}(t_k, \{f_{m,(k')}\}_{m=1}^M)$, that is, $\mathcal{V}_{(k)}$ solves the DDME to evolve $\hat{\bar{\rho}}_{[k-1]}$ to $\hat{\bar{\rho}}_{[k]}$. The costates are then written as
\begin{equation}
    \hat{\bar{\chi}}_{[k]} = \mathcal{V}_{(k+1)}^\dagger\mathcal{V}_{(k+2)}^\dagger...\mathcal{V}_{(N_{\text{T}})}^\dagger\,\hat{\rho}_{\text{targ}}.
    \label{def. DDME superoperator backwards}
\end{equation}
Similarly, we introduce the superoperator $\bar{\mathcal{U}}_{(k,k')}$ corresponding to the unitary evolution generated by the intended Hamiltonian such that
\begin{align}
    \hat{\tilde{H}}_{m,[k,k']} &\equiv \bar{\mathcal{U}}_{(k,k')}\,\hat{H}_{m} \nonumber \\
    &\equiv \, \bar{\mathcal{U}}_{(k)}\,\bar{\mathcal{U}}_{(k-1)}...\,\bar{\mathcal{U}}_{(k'+1)}\,\hat{H}_{m}\\
    &\equiv \hat{\bar{U}}_{(k)}\hat{\bar{U}}_{(k-1)}...\hat{\bar{U}}_{(k'+1)} \hat{H}_{m}\hat{\bar{U}}^\dagger_{(k'+1)}...\hat{\bar{U}}^\dagger_{(k-1)}\hat{\bar{U}}^\dagger_{(k)}.\nonumber
    \label{def. unitary superoperator}
\end{align}

The update rule that we apply to minimize $J^{(i)}$ is given by
\begin{subequations}\label{eq. Krotov update rule}
\begin{equation}
    \Delta f^{(i)}_{m,(k)} = \frac{S_{m,(k)}}{\lambda_m} \sum_{j=k}^{N_{\text{T}}} \Tr{\hat{\bar{\chi}}^{(i)}_{[j]}\frac{\partial \mathcal{K}_{(j),\{f_{m,(k')}\}}}{\partial f_{m,(k)}}\bigg{|}^{(i)}\hat{\bar{\rho}}^{(i)}_{[j]}},
    \label{eq. Krotov update rule 1}
\end{equation}
where
\begin{align}
    \begin{split}
    \frac{\partial \mathcal{K}_{(j),\{f_{m,(k')}\}}}{\partial f_{m,(k)}}\bigg{|}^{(i)}\hat{\rho} &= -\frac{i}{\hbar}\delta_{kj}[\hat{H}_m,\hat{\rho}]\\
    &-\frac{1}{\hbar^2}\sum^M_{n_1,n_2=1}\bigg{[}\hat{H}_{n_1}, \bigg{[}\frac{\partial \hat{\eta}_{n_1,n_2,(j)}}{\partial f_{m,(k)}}\bigg{|}^{(i)}, \hat{\rho}\bigg{]}\bigg{]}
    \label{eq. Krotov update rule 2}
    \end{split}
\end{align}
$\forall \hat{\rho}$ and 
\begin{align}
    \begin{split}
    \frac{\partial \hat{\eta}_{n_1,n_2,(j)}}{\partial f_{m,(k)}}\bigg{|}^{(i)} = -\frac{i(\Delta t)^2}{\hbar} \sum^{k-1}_{k'=0}&C_{n_1,n_2,(j)(k'+1)}\\
    &\bar{\mathcal{U}}^{(i)}_{[j,k]}\big{[}\hat{H}_{m}, \hat{\tilde{H}}^{(i)}_{n_2,[k,k']}\big{]}.
    \label{eq. Krotov update rule 3}
    \end{split}
\end{align}
\end{subequations}
Here $\delta_{kj}$ is the Kronecker delta. Note that the summation over future time indices in \eqref{eq. Krotov update rule 1} is present only because of the contribution from the time-nonlocal incoherent term in the DDME, and we recover the usual Krotov update step for unitary evolution if we take the correlation function to be identically 0, which is the case for unperturbed control pulses.

When Krotov's method is applied to Markovian quantum dynamics, within each iteration, each time step of all control pulses is updated sequentially from $k = 1$ to $k=N_{\text{T}}$. The quantum state must be evaluated using the updated set of control pulses from previous time steps of the current iteration, while the costates are evaluated outside the sequential update loop using control pulses from the previous iteration. The update rule can be applied to each control pulse independently. After all control pulses have been updated until $k=N_{\text{T}}$ (corresponding to the final time), the iteration number is incremented. The same process is then repeated until some predefined termination condition has been met, such as an absolute or relative tolerance on $J_{\text{T}}^{(i)} \equiv J_\text{T}(\{f_m^{(i)}(t)\}_{m=1}^M)$ or a maximum number of iterations.

For the optimization algorithm developed here, which targets at robust quantum control within the framework of disorder-dressed evolution, we maintain the general approach of Krotov's method with the termination condition defined by an absolute tolerance $J_{\text{tol}}$. However, there is one crucial difference: Since the DDME is a non-Markovian quantum master equation, the update rule for a control pulse at a specific time step depends on the disorder-averaged state in the present and all future time steps. Although it is generally possible to apply a non-Markovian quantum master equation in the Krotov framework in a time-local fashion as in \cite{hwang2012}, where an extended Liouville space was considered, here we bypass this difficulty by computing the update at time step $k$ with $\mathcal{V}_{(\kappa)}$ being fully updated $\forall\, \kappa < k$ and only partially updated $\forall\, \kappa \geq k$; that is, the superscript $(i)$ on operators (but not control pulses) in \eqref{eq. Krotov update rule} refers to evaluations based on $\{f^{(i)}_{m,(\kappa)}\}$ $\forall\, \kappa < k$ and $\{f^{(i-1)}_{m,(\kappa)}\}$ $\forall\, \kappa \geq k$. By ``partially updated" we refer to the fact that even before a control pulse gets updated at a specific time step, the propagator at this time step has already been affected by updated control pulses in the past. That is why costates are evaluated at iteration $i$ in \eqref{eq. Krotov update rule 1}, instead of at $i-1$ as in the standard Krotov method. The tradeoff here is the additional computational cost from solving the DDME over the entire future time grid in each step of the sequential update loop and the presence of the summation in \eqref{eq. Krotov update rule 1}; however, we do not focus on computational efficiency in this work. A pseudocode for the Krotov-based optimization algorithm used in this work is given in \Cref{app:1}.

\section{\label{sec:4} Single-Qubit Control Tasks}

In the following, we apply the Krotov-based DDME optimization algorithm to obtain robust control pulses for three single-qubit tasks. The three examples considered are state-to-state transfer tasks that correspond to $Z$, $X$, and Hadamard operations that are commonly applied in quantum information processing.

Throughout this section, we restrict ourselves to a single control pulse, $M = 1$, and thus abbreviate, without ambiguity, $f(t) \equiv f_1(t)$, $C(t,t') \equiv C_{1,1}(t,t')$, $S(t) \equiv S_1(t)$, and $\lambda \equiv \lambda_1$. Next we specify the drift and control Hamiltonians to be $\hat{H}_0 = \hbar\omega_0\hat{\sigma}_{z}$ and $\hat{H}_1 = \hbar\omega_0\hat{\sigma}_{x}$ for some frequency $\omega_0$ and we denote by $\hat{\sigma}_q$ the Pauli-$q$ operator for $q \in \{x, y, z\}$. Furthermore, we work in units where $\hbar = \omega_0 = 1$. To discretize time, we choose $T = 10 / \omega_0 $ and $N_{\text{T}} = 100$. We also specify the correlation function to take the stationary Gaussian form $C(t,t') = C_0\exp{-\frac{(t-t')^2}{t^2_{\text{corr}}}}$, where $t_{\text{corr}}$ is the correlation time and $C_0$ is on the order of $g_\epsilon^2(t)$. We assume $C_0 = 0.01$ and $t_{\text{corr}} = 100 / \omega_0 = 10\,T$, focusing on the limit of quasistatic pulse perturbations where robust quantum control can be maximized.

We remark that the disorder correlation strength $C_0$, which encodes the (square of the) amplitude of the pulse perturbations, is chosen such that the perturbations have a significant impact on the performance of (nonrobust) pulses, potentially reducing the purity of the disorder-averaged state at the target time by more than $20\%$ for some control tasks; nevertheless, the chosen $C_0$ is still well within the validity range of the DDME, as demonstrated by the excellent agreement between the solution of the DDME and the brute-force ensemble-averaged quantum states. Indeed, additional numerical analysis (not displayed) has shown that the approximation still works reasonably well if $C_0$ is increased by more than an order of magnitude, and the solution of the DDME may become unphysical not before $C_0>1$.

Note that, for a single qubit, our choice of $\hat{H}_0$ and $\hat{H}_1$ guarantees controllability between arbitrary (pure) initial states and (pure) target states (see, e.g., \cite{werschnik2007}). This allows us to use an initial guess pulse $h(t)$ to first obtain a Schr\"{o}dinger equation (SE)--optimized pulse $f_{\text{SE}}(t)$ that drives the initial state to the target state in the disorderless limit and then use this SE-optimized pulse as our guess pulse for the Krotov-based DDME optimizer to finally obtain the DDME-optimized pulse $f_{\text{DDME}}(t)$. We employ the standard Krotov method as used in optimal quantum control to obtain $f_{\text{SE}}(t)$ and choose $h(t)$ such that $h(0) \approx 0$ and $h(T) \approx 0$. For both types of Krotov's method, we prevent the initial and final time values of the control pulses from being updated by choosing $S(t)$ to be \cite{goerz2019}
\begin{subequations}\label{eq. choice of S(t)}
\begin{equation}
S(t) =
    \begin{cases}
        B(t;\,0, 2 \, t_{\text{on}}) & \text{for } 0 < t < t_{\text{on}} \\
        1 &  \text{for } t_{\text{on}} < t < T - t_{\text{off}} \\
        B(t;\,T - 2\,t_{\text{off}},\,T) & \text{for } T - t_{\text{off}} < t < T
    \end{cases} ,
    \label{eq. S(t) in num exp}
\end{equation}
where $B(t;t_0,t_1)$ is given by the Blackman shape \cite{harris1978}
\begin{equation}
    \begin{split}
    B(t;t_0,t_1) =&\frac{1-a}{2}-\frac{1}{2}\cos{\bigg{(}2\pi\frac{t-t_0}{t_1-t_0}\bigg{)}}\\
    &+\frac{a}{2}\cos{\bigg{(}4\pi\frac{t-t_0}{t_1-t_0}\bigg{)}}
    \label{def. blackman}
\end{split}
\end{equation}
\end{subequations}
for $a = 0.16$ and some tunable $t_{\text{on}}$ and $t_{\text{off}}$.

\begin{figure*}[htb]
	\begin{tabular}{llll}
		\includegraphics[scale=0.2]{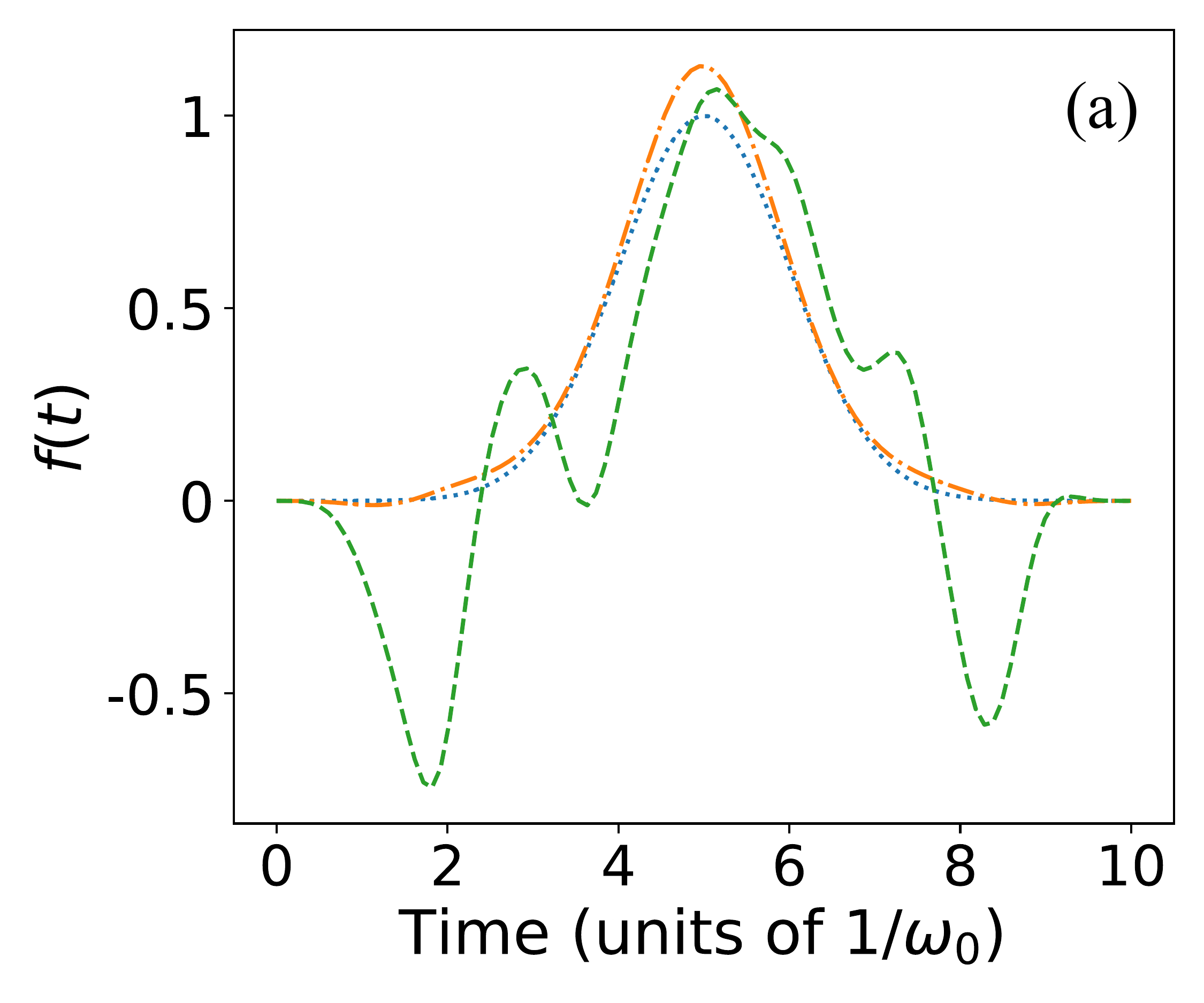}&
		\includegraphics[scale=0.204]{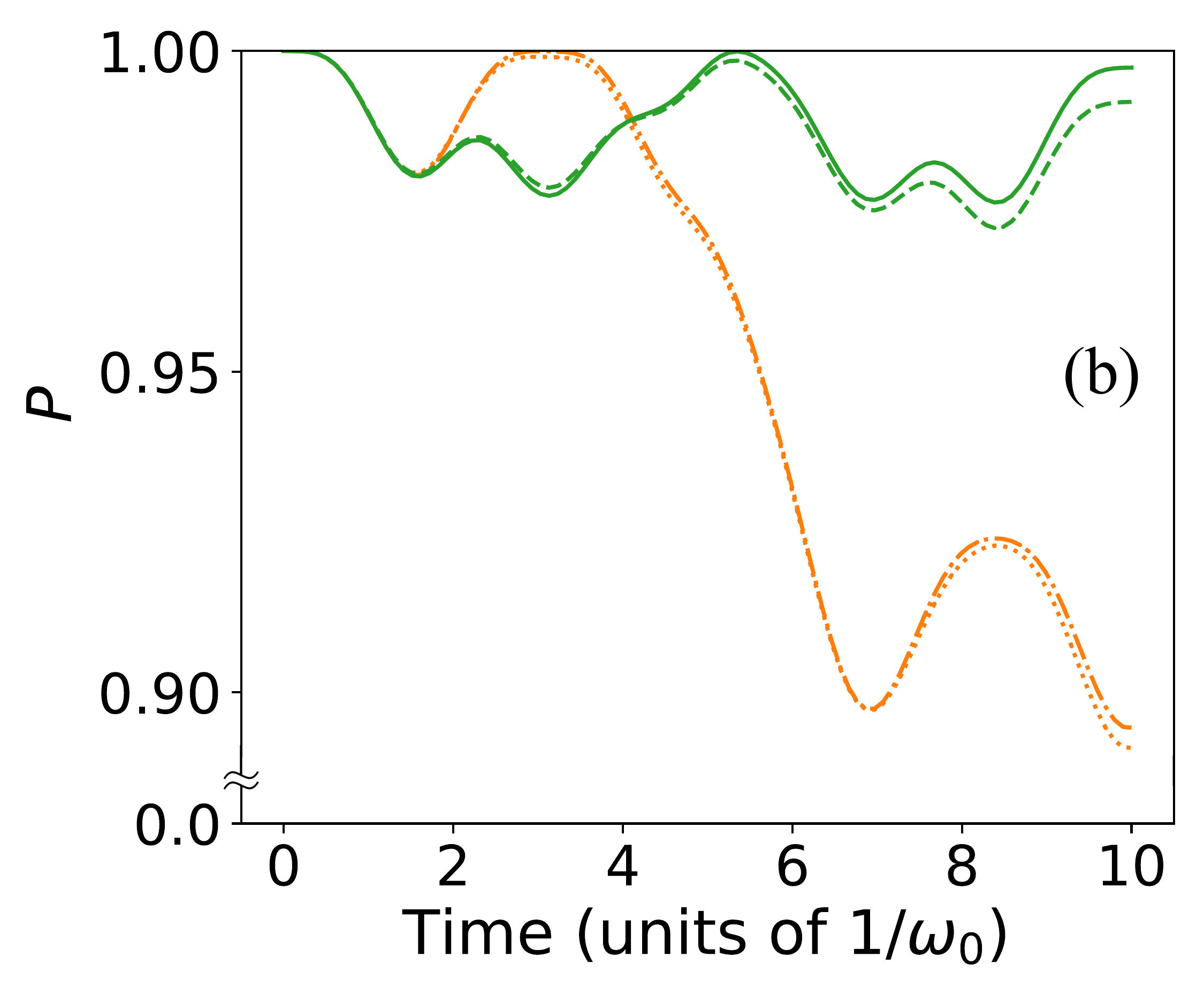}&
		\includegraphics[scale=0.2]{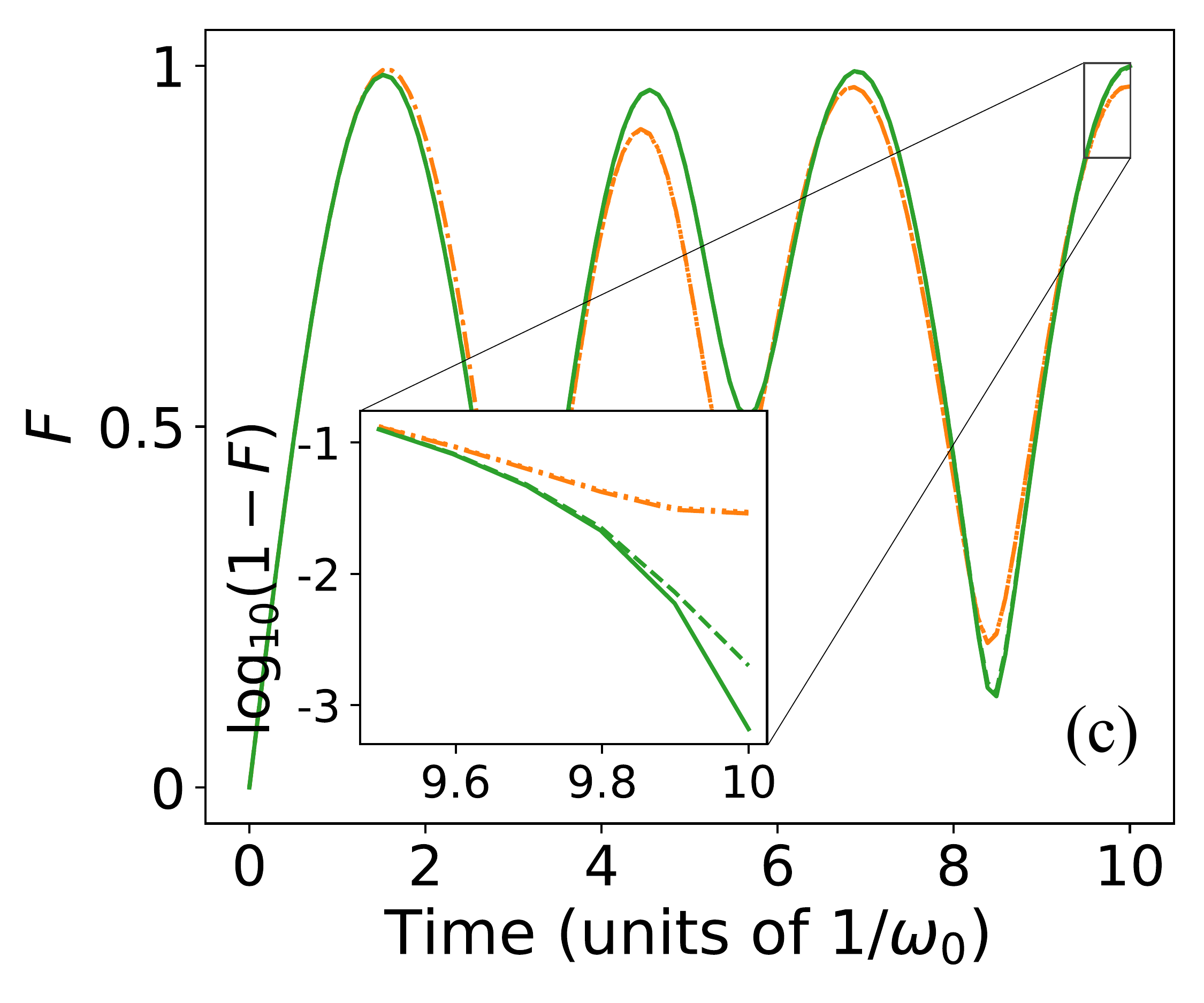}&
		\includegraphics[scale=0.2]{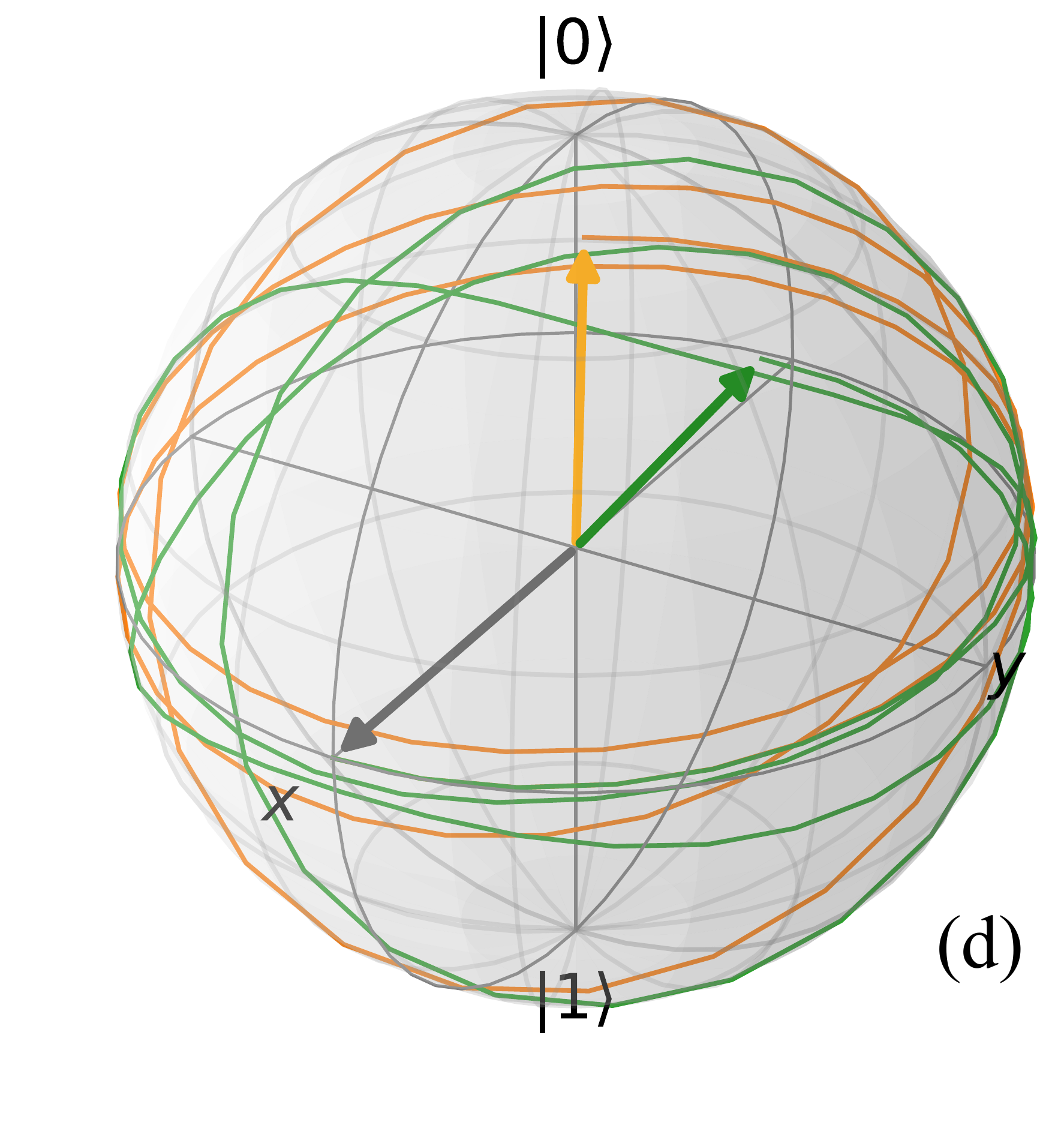}\\
		&&&\\
		\includegraphics[scale=0.2]{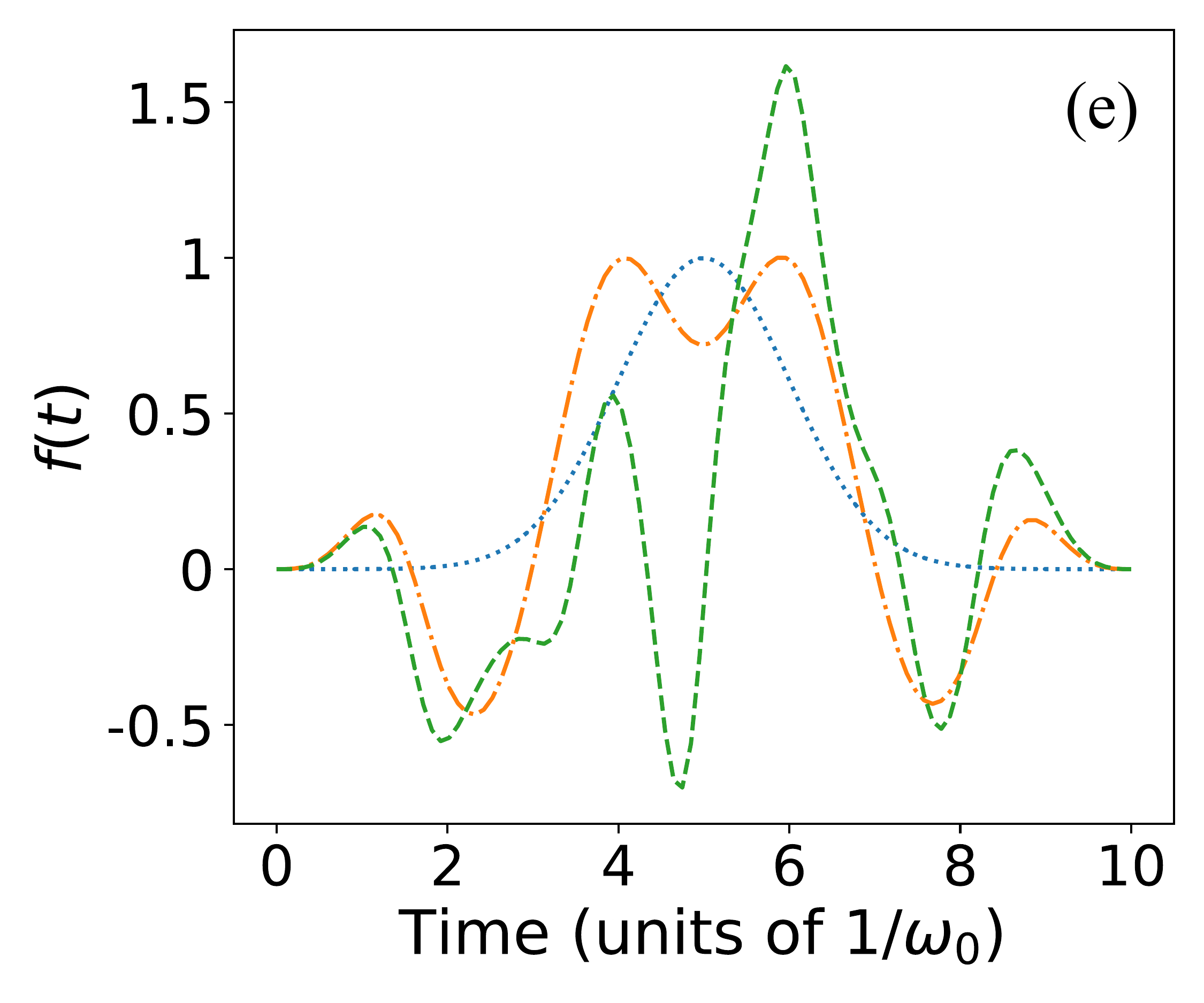}&
		\includegraphics[scale=0.204]{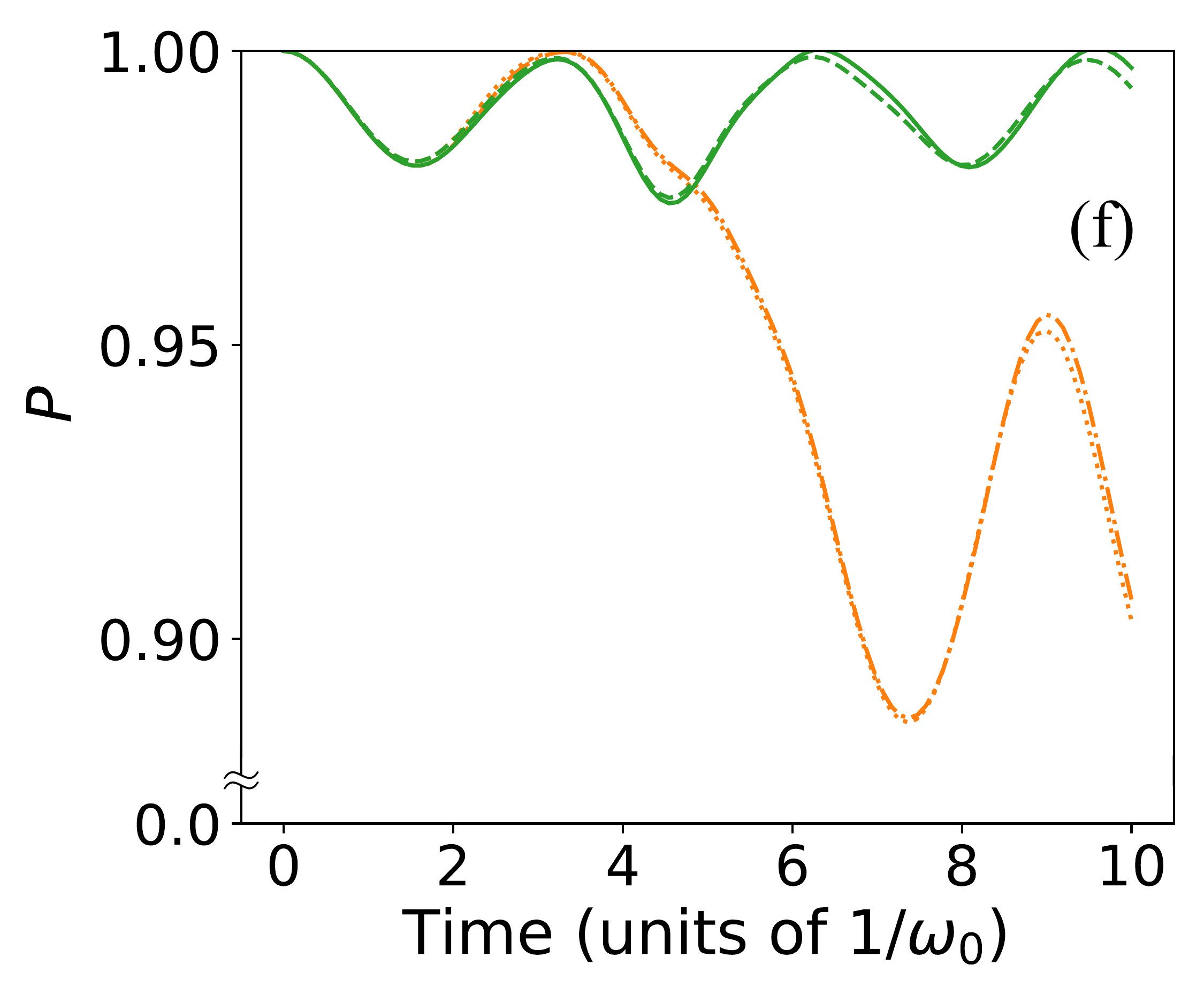}&
		\includegraphics[scale=0.2]{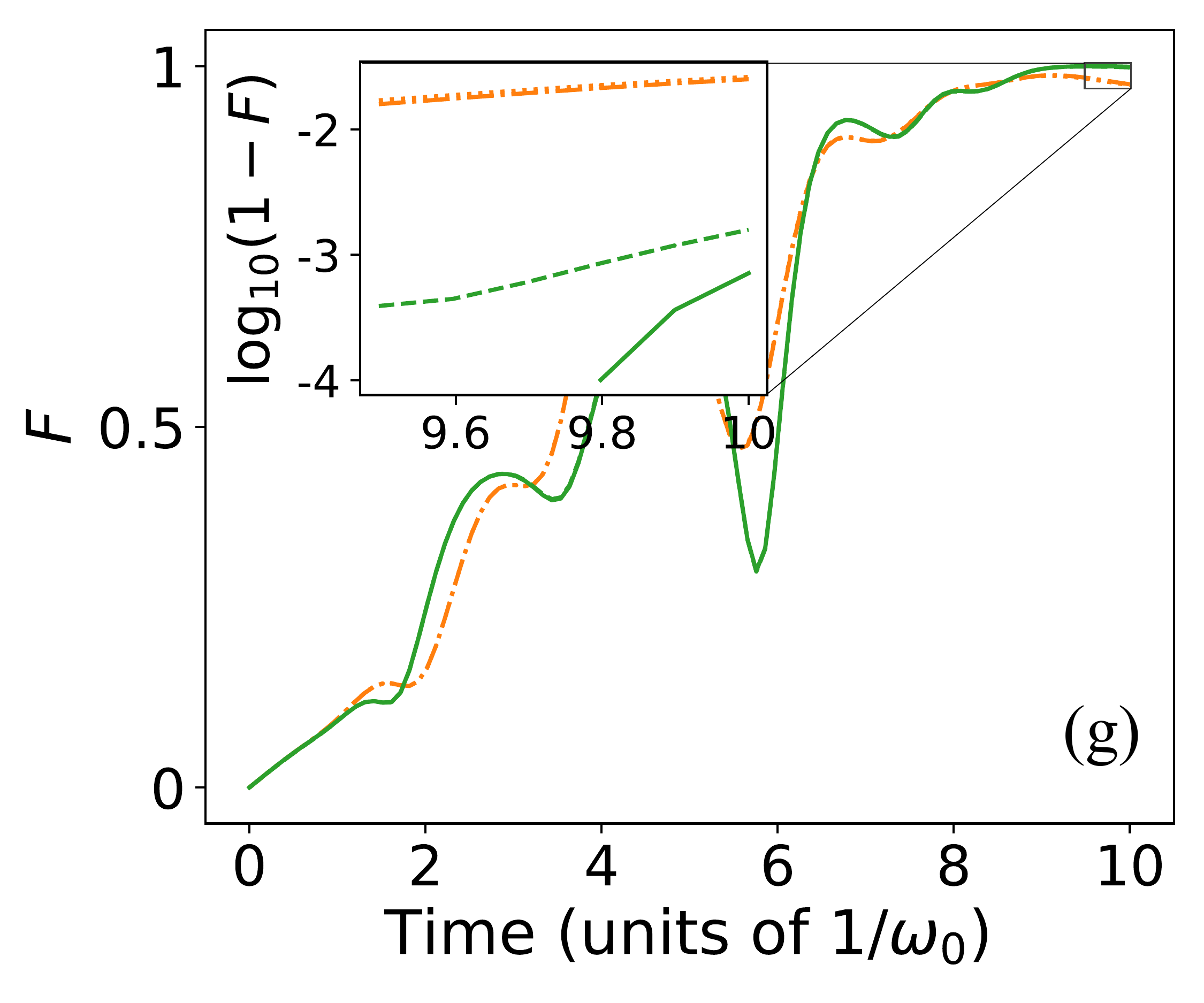}&
		\includegraphics[scale=0.2]{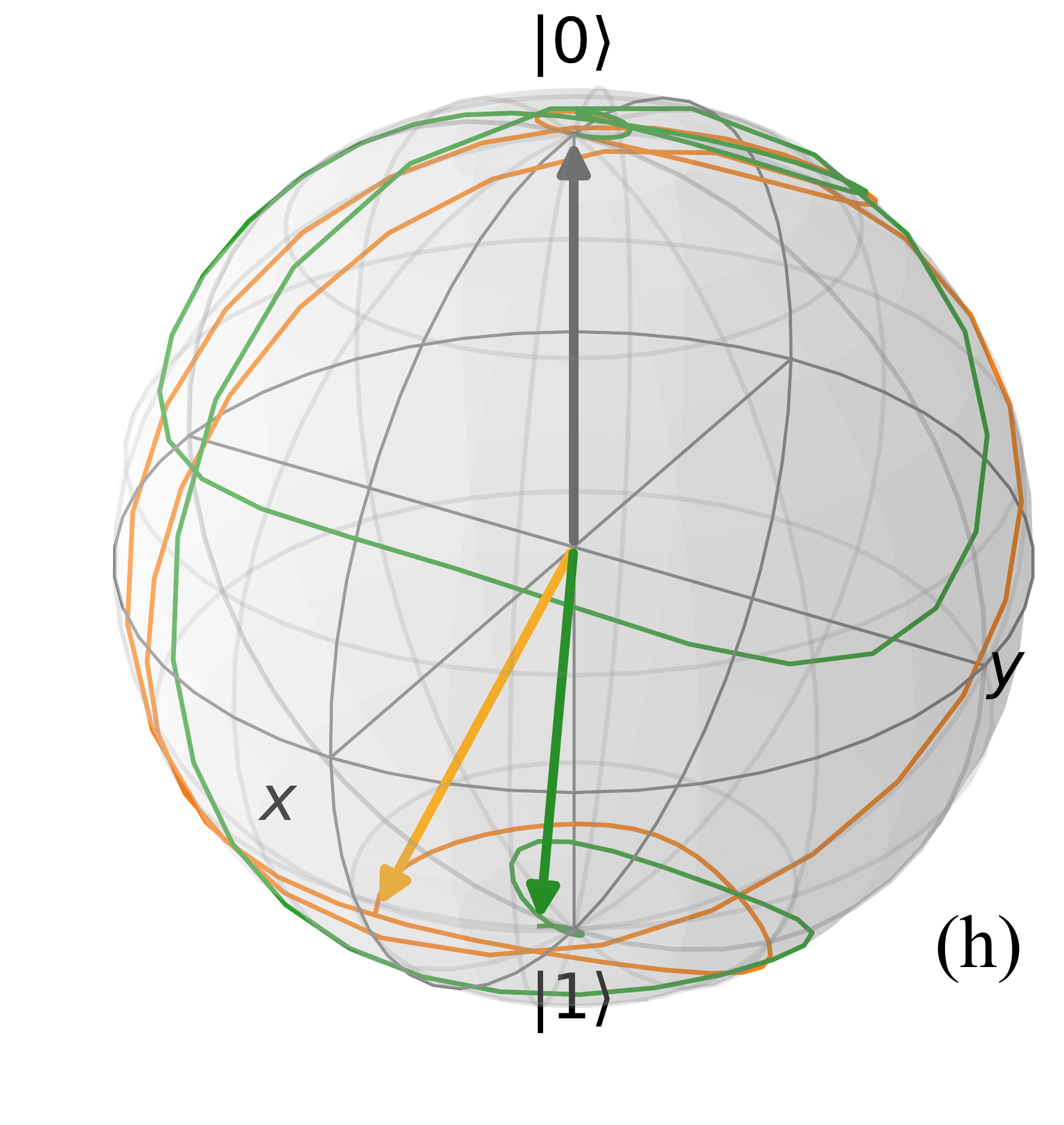}\\
		&&&\\
		\includegraphics[scale=0.2]{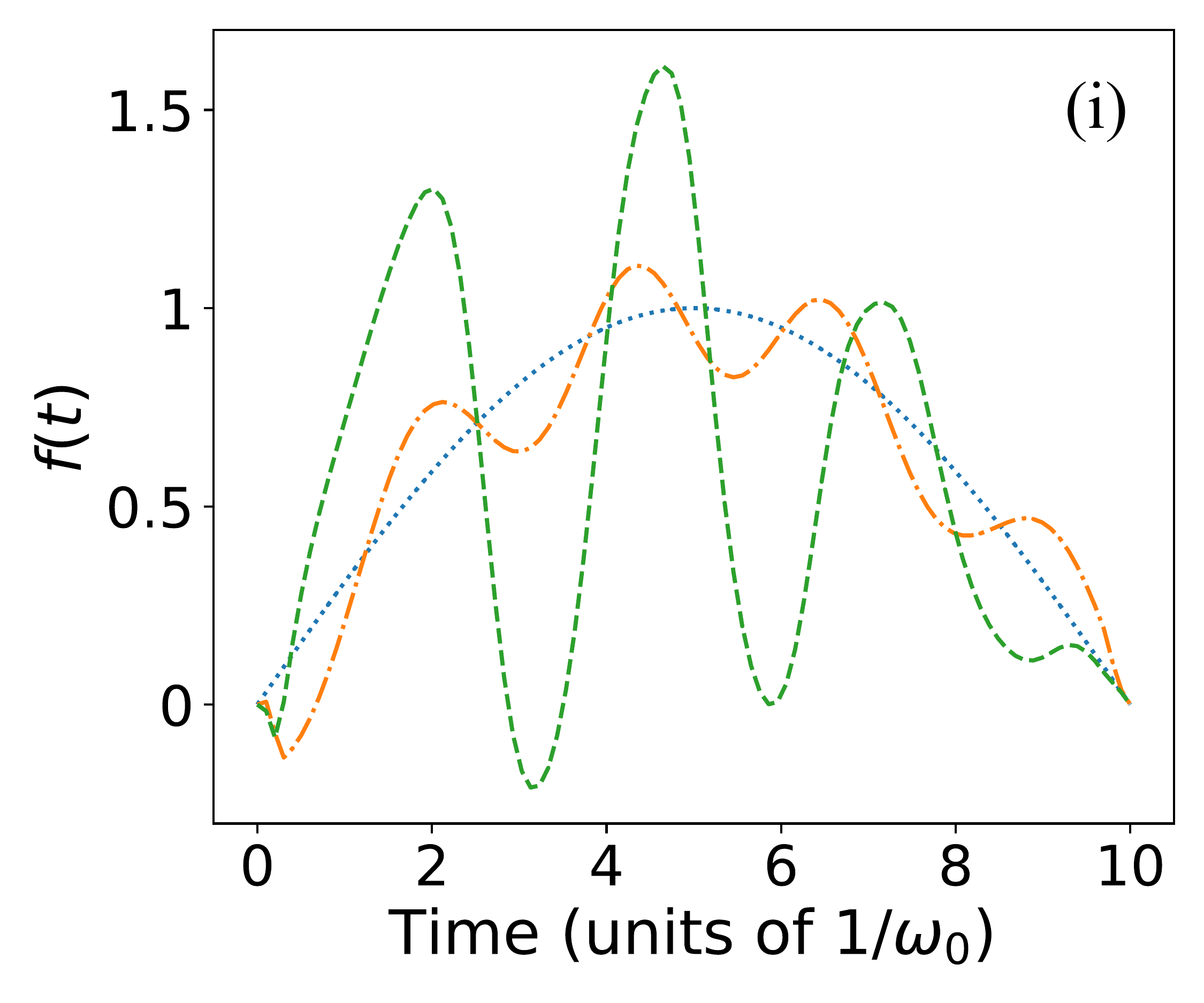}&
		\includegraphics[scale=0.204]{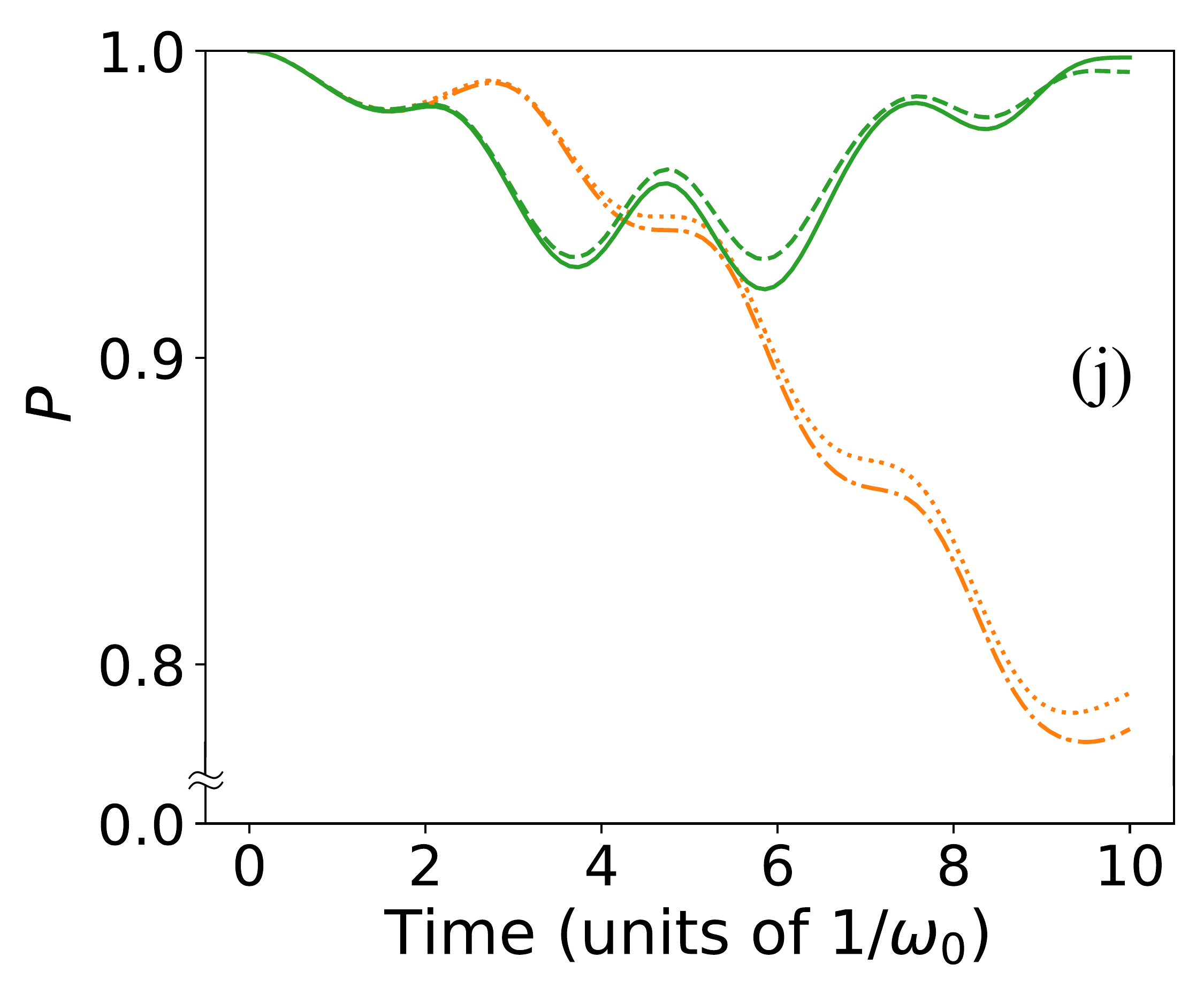}&
		\includegraphics[scale=0.2]{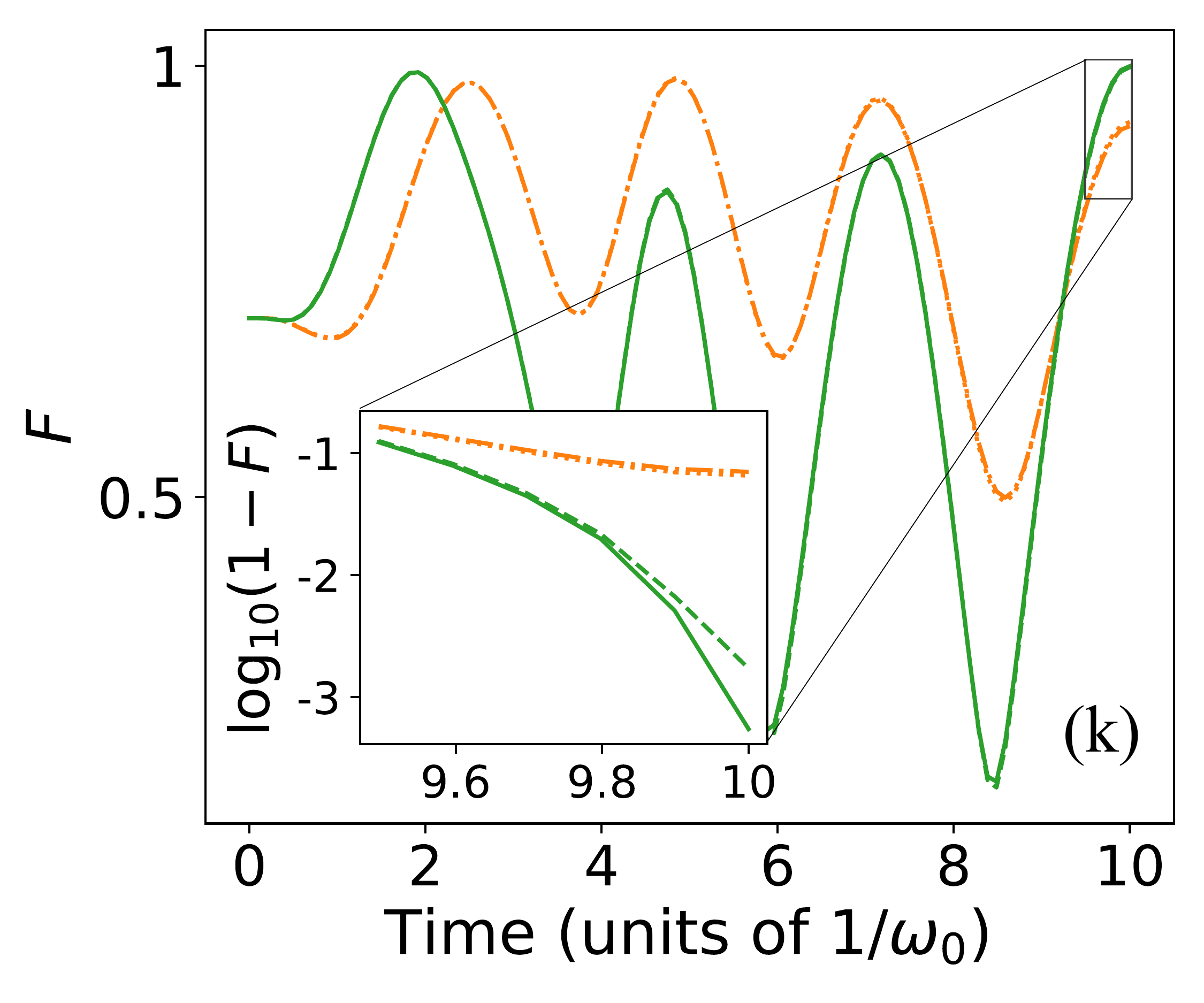}&
		\includegraphics[scale=0.2]{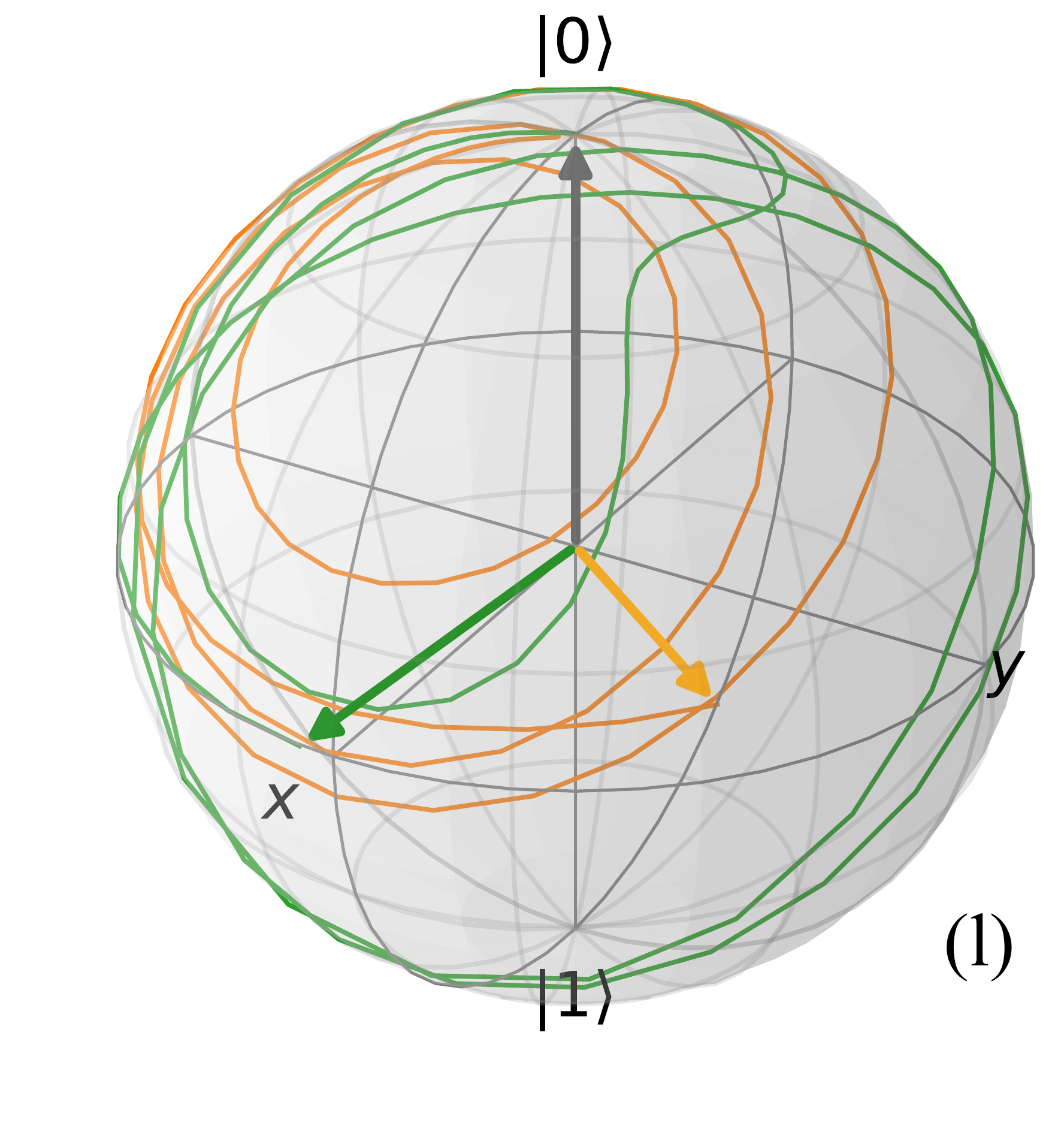}
	\end{tabular}
	\caption{Robust control for the single-qubit control tasks: (a-d) $\hat{\rho}_0 = |+\rangle \langle +|$ and $\hat{\rho}_{\text{targ}} = |-\rangle \langle -|$, (e-h) $\hat{\rho}_0 = |0\rangle \langle 0|$ and $\hat{\rho}_{\text{targ}} = |1\rangle \langle 1|$, and (i-l) $\hat{\rho}_0 = |0\rangle \langle 0|$ and $\hat{\rho}_{\text{targ}} = |+\rangle \langle +|$. (a,e,i) Compared are a DDME-optimized control pulse (dashed green line), which is informed about the statistics of pulse perturbations, and a perturbation-ignorant SE-optimized control pulse (dash-dotted orange line). The initial guess pulse (dotted blue line) is assumed to be Gaussian in (a-d) and (e-h), and $\sin(\frac{\pi t}{T})$ in (i-l). (b,f,j) If perturbations are added to the optimized pulses, the purity, $P(\hat{\bar{\rho}}(t)) = \Tr{\hat{\bar{\rho}}(t)^2}$, of the disorder-averaged state tends to decrease for the SE-optimized control pulse, while it revives for the DDME-optimized control pulse and reaches a value close to unity at the target time. The latter indicates that the differing evolutions induced by individual pulse perturbations all converge to the target state. Disorder-averaged states are obtained as solutions of the DDME (solid and dash-dotted lines) and by brute-force ensemble averaging over the evolutions induced by 4000 random pulse perturbations (dashed and dotted lines), and we find very good agreement between the two evaluation methods. This demonstrates that the DDME approximates the evolution of the disorder-averaged states well. (c,g,k) In agreement with the purity, the fidelity between the target state and the disorder-averaged state, $F(\hat{\rho}_{\text{targ}}, \hat{\bar{\rho}}(t)) = \Tr{\sqrt{\hat{\rho}_{\text{targ}}^{1/2}\hat{\bar{\rho}}(t)\hat{\rho}_{\text{targ}}^{1/2}}}$, arrives at above 0.999 for all three DDME-optimized control pulses, while it is diminished to 0.971 in (c), 0.975 in (g), and 0.930 in (k) under the SE-optimized pulses. This drastic performance discrepancy is highlighted in the insets, where the infidelities $1-F$ close to the final time are displayed on a logarithmic scale. Recall that, by construction, the SE-optimized pulses yield fidelities of unity in the absence of pulse perturbations. (d,h,l) Bloch-sphere evolution under an individual pulse perturbation. While the DDME-optimized pulse transports the initial state (dark grey arrow) close to the target state, the final state driven by the SE-optimized pulse deviates largely from the target state. This pattern holds generally throughout different disorder realizations.}\label{fig}
\end{figure*}

The first example considers $\hat{\rho}_0 = |+\rangle \langle +|$ and $\hat{\rho}_{\text{targ}} = |-\rangle \langle -|$, where $\ket{+}$ and $\ket{-}$ are the positive and negative eigenstates of $\hat{\sigma}_{x}$. Thus, the target operation corresponds to a $Z$ gate applied to a qubit initialized in the $\ket{+}$ state. We use an initial guess pulse $h(t) = \exp{-\frac{\omega_0^2(t-\frac{T}{2})^2}{2}}$, which is a Gaussian function centered at $\frac{T}{2}$. Krotov's method based on both the Schr\"{o}dinger equation and the DDME are performed with $t_{\text{on}} = t_{\text{off}} = 2 / \omega_0$, and for the latter we choose $\lambda = 1.25$ and $J_{\text{tol}} = 0.003$ to obtain $f_{\text{DDME}}(t)$.

As a second example, we investigate the case where $\hat{\rho}_0 = |0\rangle \langle 0|$ and $\hat{\rho}_{\text{targ}} = |1\rangle \langle 1|$, so that the target operation corresponds to an $X$ gate applied to a qubit initialized in the $\ket{0}$ state. We continue to use the same $h(t)$ and $S(t)$ as in the previous example to obtain $f_{\text{SE}}(t)$; however, this time we choose $\lambda = 0.5$ and $J_{\text{tol}} = 0.003$ to obtain $f_{\text{DDME}}(t)$ with a higher learning rate.

Finally, we consider the transition from $\hat{\rho}_0 = |0\rangle \langle 0|$ to $\hat{\rho}_{\text{targ}} = |+\rangle \langle +|$ so that the target operation corresponds to a Hadamard gate applied to a qubit initialized in the $\ket{0}$ state. For this example, we choose $h(t) = \sin{(\frac{\pi t}{T})}$ and $S(t)$ with $t_{\text{on}} = t_{\text{off}} = 0.3 / \omega_0$. Here $f_{\text{DDME}}(t)$ is then obtained from $f_{\text{SE}}(t)$ with $\lambda = 1.25$ and $J_{\text{tol}} = 0.003$.

The results of the numerical experiments for the three examples are shown in \Cref{fig} (a-d), (e-h), and (i-l) in the same order, where each plot in the same vertical line displays the same features across the different examples. Curves associated with $f_{\text{SE}}(t)$ are shown in orange, while those associated with $f_{\text{DDME}}(t)$ are colored in green. For each of the examples, we show $h(t)$ (blue dotted), $f_{\text{SE}}(t)$ (orange dash-dotted), and $f_{\text{DDME}}(t)$ (green dashed) in \Cref{fig} (a,e,i).

To compare the performance of the SE-optimized and the DDME-optimized control pulses with respect to robustness, we solve the disorder-dressed evolution for both control pulses and compare the resulting state purities. In particular, a final-time purity close to (or of exactly) unity indicates that the state trajectories associated with different disorder realizations have all arrived close to (or exactly at) the target state.

The results of these purity comparisons are shown in \Cref{fig} (b,f,j). To demonstrate the excellent approximation of the DDME, we determine the disorder-dressed evolution in two ways: by solving the DDME (solid and dash-dotted lines) and by numerically exact brute-force averaging (dashed and dotted lines) as described by the definition \eqref{def. disorder-averaged state} of the disorder-averaged quantum state. In the latter case, we average over 4000 random realizations of symmetric Gaussian noises $g_\epsilon(t)$ according to a Gaussian probability distribution and in agreement with the correlation function $C(t,t')$. We find very good agreement between the two methods within the timescale considered.

Consistently across the examples, we observe that, while the state purity under the SE-optimized evolution exhibits an overall decreasing trend, the state purity under the DDME-optimized evolution recovers after some time and rises close to unity at the final time. Thus, we observe that, as expected, $f_{\text{DDME}}(t)$ exhibits significantly increased robustness against disorder. 

In \Cref{fig} (c,g,k) we display the fidelities between the disorder-averaged state and the target state for both the evolution generated by the SE-optimized (orange dash-dotted and dotted lines) control pulse and the evolution generated by the DDME-optimized (green solid and dashed lines) control pulse, where the disorder-dressed evolutions are again obtained both by solving the DDME and by brute-force averaging. To highlight the most relevant region, we magnify the final-time infidelities in the insets on a logarithmic scale. Consistent with the purity evolutions, the DDME-optimized pulses achieve final-time fidelities above $0.999$ for all examples, while about $3\%$ in (c), $2\%$ in (g), and $7\%$ in (k) are lost with the SE-optimized pulses. This strikingly demonstrates the robustness boost that is obtained with the disorder-dressed evolution approach.

Finally, for concreteness, we show in \Cref{fig} (d,h,l) the Bloch-sphere trajectories for a single arbitrarily chosen disorder realization when the qubit is driven by either the SE-optimized (orange) or the DDME-optimized (green) control pulse. For each example, we observe that the final state under the SE-optimized evolution deviates largely from the target state, while the final state of the DDME-optimized evolution remains close to the target state. This pattern holds for other disorder realizations as well and further confirms the robustness of the DDME-optimized pulse.

\begin{figure}[t]
  \centering
  \includegraphics[width=\columnwidth]{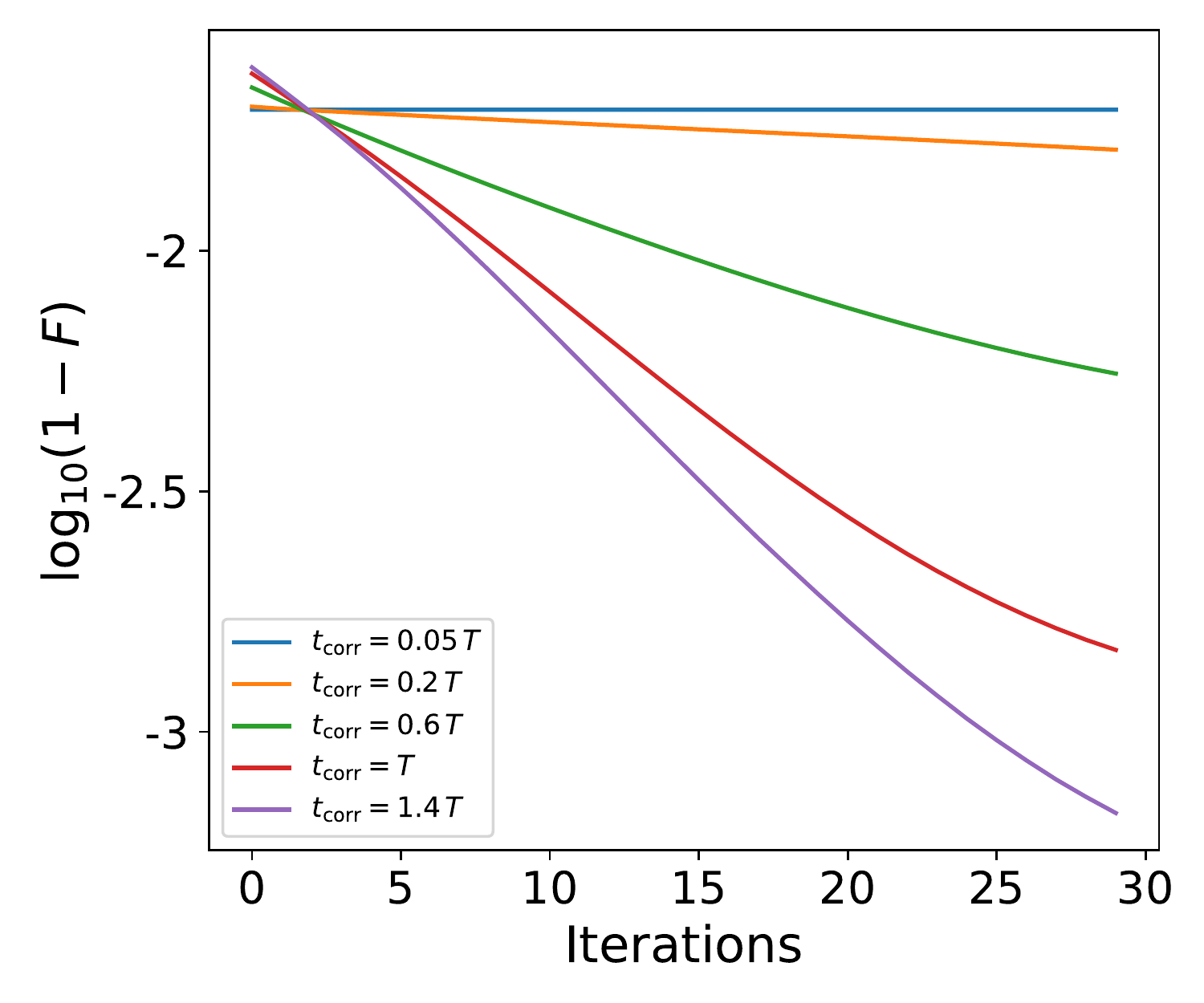}
  \caption{Role of temporal correlations in the pulse perturbations for the prospect of robust control. For the control task $\hat{\rho}_0 = |+\rangle \langle +|$ and $\hat{\rho}_{\text{targ}} = |-\rangle \langle -|$ (the first example discussed in \Cref{sec:4}), and starting from the SE-optimized control pulse, the Krotov-based optimization is performed for different correlation times $t_{\text{corr}} \in \{0.05\,T, 0.2\,T, 0.6\,T, T, 1.4\,T\}$ (displayed as lines from top to bottom corresponding to ascending order of $t_{\mathrm{corr}}$) for $T = 10/\omega_0$ while fixing all other parameters. Without specifying an absolute tolerance, the optimizer is run for 30 iterations. We find that the infidelity does not decrease in the near-Markovian limit ($t_{\text{corr}} = 0.05\,T$), indicating that robust control becomes impossible in this limit, in agreement with \eqref{eq. Gaussian white noise DDME in Lindblad form}. With increasing correlation time, we observe a monotonic cross-over to the quasi-static case ($t_{\text{corr}} = 1.4\,T$), where the convergence speed of the algorithm and the fidelity of the disorder-averaged state with the target state increase with increasing correlation time. The performance difference in terms of the infidelity reduction can, depending on the correlation time, span several orders of magnitude.}
  \label{fig2}
\end{figure}

Let us repeat that we have focused on quasistatic pulse perturbations ($t_{\rm corr} \gg T$), since in this limit the performance of robust quantum control can be maximized and full purity revivals can in principle be achieved, as exposed by our numerical examples. In contrast, the opposite limit of vanishing temporal correlations severely limits robust control~[cf.~\eqref{eq. Gaussian white noise DDME in Lindblad form}]. We also verified this numerically with $t_{\rm corr} = 0.05\,T$ in \Cref{fig2}, where our algorithm was not able to deliver fidelity increases when starting with SE-optimized pulses. In between these two extreme cases, we observe a monotonic crossover, where the convergence speed of the algorithm, the maximum achievable purity of the disorder-averaged state, and the fidelity of the disorder-averaged state with the target state decrease with decreasing correlation time (see~\Cref{fig2}).

\section{\label{sec:5} Conclusions}

We have demonstrated how robust control pulses can be systematically identified with the help of disorder-dressed evolution equations. The latter apply in the perturbative limit of weak pulse distortions. In contrast to schemes based on searches over random ensembles, our approach is deterministic, relying on the maximization of the purity of the disorder-averaged state. We expect that this conceptually founded approach will further deepen our understanding of what constitutes robust control pulses, and in special cases analytical solutions may be possible. For the automatized numerical determination of robust control pulses in field applications, we have developed an adapted and generalized variant of Krotov's method. Our single-qubit demonstrations expose the power of our method, indicated by target-state fidelities beyond 0.999, which amounts to improvements of up to two orders of magnitude across the examples.

To formulate the underlying disorder-dressed evolution equation, we have generalized existing formulations to time-dependent Hamiltonians; moreover, we have adapted them to (in general time-dependent) pulse perturbations. In our numerical analysis, we focused on the (quasistatic) limit of correlation times larger than the pulse duration, where pulse perturbations vary slowly over the temporal extent of the pulse. In this limit, the disorder-dressed evolution becomes highly non-Markovian and (in principle full) purity revivals can emerge.

We have adopted Krotov's method for our numerical implementation, and its successful application to several single-qubit control tasks verifies the viability of the algorithm. Irrespectively, the main focus of this work is conceptual, and the adoption of other optimal control algorithms to the disorder-dressed evolution may yield further performance improvements. Moreover, a comparison of the computational complexity of the disorder-dressed approach with the computational complexities of other approaches to robust control may be insightful. While there is an increased cost per iteration due to the adaption of the disorder-dressed master equation to the updated pulse at each time step, our numerical experiments indicate that the required number of iterations may be reduced by several orders of magnitude compared to, e.g., ensemble optimization. For the single-qubit tasks considered above our algorithm converges after fewer than $30$ iterations.

While we restricted our numerical analysis to proof-of-principle demonstrations with single qubits and single control pulses, our method and the developed algorithm are applicable to general (finite-dimensional) quantum systems and arbitrary numbers of control pulses. For example, a natural next step would be to address the robust control of entangling two-qubit gates. Moreover, the DDME formalism is easily adapted to error sources other than pulse perturbations, such as, e.g., disorder on the drift Hamiltonian. Finally, while the presented formalism is designed for the mitigation of coherent error sources (i.e., disorder in the Hamiltonian), it should be clear that the formalism and code can be naturally extended to include also decoherence channels induced by environmental coupling. These channels would then, to first order in the sufficiently small environment-induced decoherence rates, be added as (Markovian) incoherent dynamical terms to the evolution of the disorder-averaged quantum state.

\section*{Acknowledgments}

Part of the code used in the numerical experiments utilizes the tools provided by QuTiP \cite{johansson2012, johansson2013}.
C.G. would like to thank D. Burgarth for discussions during his visits, partly funded by the Australian Research Council, Project No. FT190100106.
F.N. was supported in part by
Nippon Telegraph and Telephone Corporation (NTT) Research,
the Japan Science and Technology Agency (JST) [via
the Quantum Leap Flagship Program (Q-LEAP),
Moonshot R\&D Grant No. JPMJMS2061],
the Japan Society for the Promotion of Science (JSPS)
[via Grants-in-Aid for Scientific Research (KAKENHI) Grant No. JP20H00134],
the Army Research Office (ARO) (Grant No. W911NF-18-1-0358),
the Asian Office of Aerospace Research and Development (AOARD) (via Grant No. FA2386-20-1-4069), and
the Foundational Questions Institute Fund (FQXi) via Grant No. FQXi-IAF19-06.

\onecolumngrid

\begin{algorithm}[H]
	\caption{Krotov-based Optimization Algorithm for Robust Quantum Control}\label{algorithm 1}
	\textbf{Inputs and auxiliary functions:}
	\begin{enumerate}
	    \item Initial density matrix \tab $\hat{\rho}_0$
	    \item Target density matrix \tab $\hat{\rho}_{\text{targ}}$
	    \item Drift Hamiltonian \tab $\hat{H}_0$
        \item Control Hamiltonians \tab $\{\hat{H}_m\}_{m=1}^M$
        \item Guess pulses \tab $\{\{f_{m,(k)}^{\text{guess}}\}_{m=1}^M\}_{k=1}^{N_{\text{T}}}$
        \item Correlation functions \tab $\{\{C_{n_1,n_2,(j)(l)}\}_{n_1,n_2 = 1}^M\}_{j,l=1}^{N_{\text{T}}}$
        \item Update shape functions \tab $\{\{S_{m,(k)}\}_{m=1}^M\}_{k=1}^{N_{\text{T}}}$
        \item Inverse Krotov step sizes \tab $\{\lambda_m\}_{m=1}^M$
        \item Absolute cost tolerance \tab $J_{\text{T}}^{\text{tol}}$
        \item Maximum number of iterations \tab $i_{\text{max}}$
    	\item Unitary Solver \tab $\bar{\mathcal{U}}^{(i)}_{[j,j']}(\;\cdot\;)$
        \item DDME Solver \tab $\mathcal{V}^{(i)}_{[j,j']}(A$;$\;\cdot\;)$
        \item Backward DDME Solver \tab $\mathcal{V}^{\dagger(i)}_{[j,j']}(A$;$\;\cdot\;)$
    \end{enumerate}
	\textbf{Success Criterion}: $\exists$ iteration number $i$ such that $i \leq i_{\text{max}}$ and $J^{(i)}_{\text{T}} \equiv J_{\text{T}}(\{\{f_{m,(k)}^{(i)}\}_{m=1}^M\}_{k=1}^{N_{\text{T}}}) \leq J_{\text{T}}^{\text{tol}}$. Failure otherwise.\\
	\textbf{Output:} Optimized set of control pulses $\{\{f_{m,(k)}^{\text{opt}}\}_{m=1}^M\}_{k=1}^{N_{\text{T}}}$ such that $J^{(i)}_{\text{T}} \leq J_{\text{T}}^{\text{tol}}$.
    \begin{algorithmic}[1]
        \Procedure{DDME\_Krotov\_Optimization}{$\hat{\rho}_0$, $\hat{\rho}_{\text{targ}}$, $\hat{H}_0$, $\{\hat{H}_m\}$, $\{f_{m,(k)}^{\text{guess}}\}$, $\{C_{n_1,n_2,(j)(l)}\}$, $\{S_{m,(k)}\}$, $\{\lambda_m\}$, $J_{\text{T}}^{\text{tol}}$, $i_{\text{max}}$}
		\State allocate storage array $\Phi[0\ldots N_{\text{T}}]$ \algorithmiccomment{for $\hat{\bar{\rho}}(t)$}
		\State allocate storage array $X[0\ldots N_{\text{T}}]$ \algorithmiccomment{for $\hat{\bar{\chi}}(t)$}
		\State allocate storage array $A[1\ldots M, 1\ldots M, 0\ldots N_{\text{T}}]$ \algorithmiccomment{for $\hat{\eta}_{n_1,n_2}(t)$}
		\State allocate storage array $B[1\ldots M, 0\ldots N_{\text{T}}, 0\ldots N_{\text{T}}]$ \algorithmiccomment{for $\hat{\tilde{H}}_m(t,t')$}
		\State $\Phi[0] \leftarrow \hat{\rho}_0$
		\State $X[N_{\text{T}}] \leftarrow \hat{\rho}_{\text{targ}}$
		\State $\forall m,k$ : $f_{m,(k)}^{(0)}\leftarrow f_{m,(k)}^{\text{guess}}$ \algorithmiccomment{initial guess pulse}
		\State $B \leftarrow$
	        \textsc{$\hat{\tilde{H}}^{(0)}$\_Solver}($\,\cdots$; $B$)
		\State $A \leftarrow$ \textsc{$\hat{\eta}^{(0)}$\_Solver}($\,\cdots$; $B, A$)
		\State $\Phi[N_{\text{T}}] \leftarrow \mathcal{V}^{(0)}_{[N_{\text{T}},0]}(A; \Phi[0])$
		\State $J^{(0)}_{\text{T}} \leftarrow 1-\Tr{X[N_{\text{T}}]\Phi[N_{\text{T}}]}$ \algorithmiccomment{cost before optimization \eqref{def. J_T}}
		\State $i \leftarrow 0$  \algorithmiccomment{iteration number}
		\While {$J^{(i)}_{\text{T}} > J_{\text{T}}^{\text{tol}}$ and $i < i_{\text{max}}$} \algorithmiccomment{optimization loop}
			\State $i \leftarrow i + 1$
			\State $\forall m,k$ : $f_{m,(k)}^{(i)}\leftarrow f_{m,(k)}^{(i-1)}$
		    \For {$k \leftarrow 1,2,\ldots,N_{\text{T}}$}
		    \algorithmiccomment{sequential update loop}
		    \If{$k \neq N_{\text{T}}$}
    		    \For {$j \leftarrow N_{\text{T}}-1, N_{\text{T}}-2, \ldots, k$}
                    \State $X[j] \leftarrow \mathcal{V}^{\dagger(i)}_{[j,j+1]}(A; X[j+1])$ \algorithmiccomment{store $\hat{\bar{\chi}}^{(i-1)}(t) \; \forall $ future time steps}
                \EndFor
            \EndIf
	        \For {$j \leftarrow k,k+1,\ldots,N_{\text{T}}$}
	            \State $\Phi[j] \leftarrow \mathcal{V}^{(i)}_{[j,j-1]}(A; \Phi[j-1])$ \algorithmiccomment{store $\hat{\bar{\rho}}^{(i)}(t) \; \forall $ future time steps}
	        \EndFor
		    \For {$m \leftarrow 1,2,\ldots,M$} \algorithmiccomment{update each control pulse independently}
		        \State $D_{m,(k)} \leftarrow$ \textsc{$D^{(i)}$\_Solver}($\,\cdots$; $\Phi$, $X$, $B$, $m$, $k$) \algorithmiccomment{obtain gradient \eqref{eq. Krotov update rule}}
	            \State $f^{(i)}_{m,(k)} \leftarrow f^{(i-1)}_{m,(k)} + \frac{S_{m,(k)}}{\lambda_m}D_{m,(k)}$ \algorithmiccomment{apply update \eqref{eq. Krotov update rule 1}}
			\EndFor
			\State $B \leftarrow$
	        \textsc{$\hat{\tilde{H}}^{(i)}$\_Solver}($\,\cdots$; $B$)
	        \algorithmiccomment{recalculating $A$ \& $B$ after sequential update step}
	        \State $A \leftarrow$
	        \textsc{$\hat{\eta}^{(i)}$\_Solver}($\,\cdots$; $B, A$)
			\State $\Phi[k] \leftarrow \mathcal{V}^{(i)}_{[k,k-1]}(A; \Phi[k-1])$ \algorithmiccomment{replace $\Phi[k]$ with the one evolved with updated $\{f_{m,(k)}\}_{m=1}^M$}
			\EndFor
			\State $J^{(i)}_{\text{T}} \leftarrow 1-\Tr{X[N_{\text{T}}]\Phi[N_{\text{T}}]}$ \algorithmiccomment{obtain cost after iteration $i$ \eqref{def. J_T}}
		\EndWhile
		\If{$J^{(i)}_{\text{T}} \leq J^{\text{tol}}_{\text{T}}$} 
		    \State $\forall m,k$ : $f_{m,(k)}^{\text{opt}}\leftarrow f_{m,(k)}^{(i)}$
		    \State \Return $\{\{f_{m,(k)}^{\text{opt}}\}_{m=1}^M\}_{k=1}^{N_{\text{T}}}$ \algorithmiccomment{return optimized set of control pulses if converged}
		 \EndIf
		 \EndProcedure
		\algstore{part1}
    \end{algorithmic}
\end{algorithm}
\begin{algorithm}[H]
    \ContinuedFloat
    \caption{Krotov-based Optimization Algorithm for Robust Quantum Control (continued)}
    \begin{algorithmic}[1]
        \algrestore{part1}
        \Procedure{$D^{(i)}$\_Solver}{$\hat{H}_0$, $\{\hat{H}_m\}$, $\{f_{m,(k)}^{(i)}\}$, $\{C_{n_1,n_2,(j)(l)}\}$; $\Phi$, $X$, $B$, $m$, $k$}
		\State $\tilde{D}_1 \leftarrow - \frac{i}{\hbar}[\hat{H}_m,\Phi[k]]$ \algorithmiccomment{derivative of coherent term}
        \State $D \leftarrow \Tr{X[k]\tilde{D}_1}$
        \For {$j \leftarrow k,k+1\ldots,N_{\text{T}}$} \algorithmiccomment{summation from product rule \eqref{eq. Krotov update rule 1}}
        \State $\tilde{D}_2 \leftarrow \mathbf{0}$
        \For {$n_1 \leftarrow 1,2\ldots,M$}
        \For {$n_2 \leftarrow 1,2\ldots,M$}
        \State $\tilde{D}'_2 \leftarrow \mathbf{0}$
        \For {$l \leftarrow 0,1\ldots,k-1$} \algorithmiccomment{Riemann sum \eqref{eq. Krotov update rule 3}}
            \State $\tilde{D}'_2 \leftarrow \tilde{D}'_2 +  C_{n_1,n_2,(j)(l+1)}\bar{\mathcal{U}}^{(i)}_{[j,k]}([\hat{H}_{m}, B[n_2,k,l]])$
        \EndFor
        \State $\tilde{D}'_2 \leftarrow -\frac{i(\Delta t)^2}{\hbar}\tilde{D}'_2$
        \State $\tilde{D}_2 \leftarrow \tilde{D}_2 + [\hat{H}_{n_1},[\tilde{D}'_2,\Phi[j]]]$
        \EndFor
        \EndFor
        \State $\tilde{D}_2 \leftarrow -\frac{1}{\hbar^2}\tilde{D}_2$
        \State $D \leftarrow D + \Tr{X[j]\tilde{D}_2}$ \algorithmiccomment{add derivative of incoherent terms}
        \EndFor
        \State \Return $D$ \algorithmiccomment{derivative of coherent \& incoherent terms}
        \EndProcedure
        \item[]
		\Procedure{$\hat{\eta}^{(i)}$\_Solver}{$\{C_{n_1,n_2,(j)(l)}\}$; $B$, $A$}
		\For {$n_1 \leftarrow 1,2,\ldots,M$}
		    \For {$n_2 \leftarrow 1,2,\ldots,M$}
		        \State $A[n_1,n_2,0] \leftarrow \mathbf{0}$
		        \For {$j \leftarrow 1,2,\ldots,N_{\text{T}}$}
		            \State $\tilde{A} \leftarrow \mathbf{0}$
		            \For {$l \leftarrow 0,1,\ldots,j-1$} \algorithmiccomment{Riemann sum \eqref{def. eta}}
		                \State $\tilde{A} \leftarrow \tilde{A} + \Delta t \,C_{n_1,n_2,(j)(l+1)}B[n_2,j,l]$
		            \EndFor
                    \State $A[n_1,n_2,j] \leftarrow \tilde{A}$
                \EndFor
		    \EndFor
		\EndFor
        \State \Return $A$
        \EndProcedure
		\item[]
		\Procedure{$\hat{\tilde{H}}^{(i)}$\_Solver}{$\hat{H}_0$, $\{\hat{H}_m\}$, $\{f_{m,(k)}^{(i)}\}$; $B$}
		\For {$m \leftarrow 1,2,\ldots,M$}
    		\For {$j \leftarrow 0,1,\ldots,N_{\text{T}}$}
                \For {$l \leftarrow 0,1,\ldots,j$}
                    \State $B[m, j, l] \leftarrow \bar{\mathcal{U}}^{(i)}_{[j,l]}(\hat{H}_m)$ \algorithmiccomment{coherently evolve each $\hat{H}_m$}
                \EndFor
            \EndFor
        \EndFor
        \State \Return $B$
        \EndProcedure
	\end{algorithmic}
\end{algorithm}

\twocolumngrid

\appendix                                     
\section{\label{app:1}Pseudocode for Krotov-based Optimization Algorithm}

We present in \Cref{algorithm 1} the pseudocode for the Krotov-based optimization algorithm for robust quantum control introduced in the main text, following an implementation inspired by \cite{goerz2019}, but highly modified. The pseudocode terminates with the satisfaction of an absolute tolerance $J_{\text{T}}^{\text{tol}}$ or a maximum number of iteration $i_{\text{max}}$, where if there exists $i \leq i_{\text{max}}$ such that $J^{(i)} < J_{\text{T}}^{\text{tol}}$, then the algorithm succeeds and outputs a set of discretized optimal control pulses $\{\{f_{m,(k)}^{\text{opt}}\}_{m=1}^M\}_{k=1}^{N_{\text{T}}}$. Otherwise, the algorithm fails, and terminates right after the iteration where $i = i_{\text{max}}$. We take the unitary [generated by $\hat{\bar{H}}(t)$], DDME, and backward DDME solvers to be given functions, and denote their evolutions from $t=t_{j'}$ to $t=t_j$ for $j \geq j'$ by $\bar{\mathcal{U}}^{(i)}_{[j,j']}(\;\cdot\;)$, $\mathcal{V}^{(i)}_{[j,j']}(A$;$\;\cdot\;)$, and $\mathcal{V}^{\dagger(i)}_{[j,j']}(A$;$\;\cdot\;)$, respectively. Here, the unitary solver depends on $\hat{H}_0$, $\{\hat{H}_m\}_{m=1}^M$ and $\{\{f_{m,(k)}^{(i)}\}_{m=1}^M\}_{k=j'+1}^{j}$, but we suppress these dependences in the pseudocode for clarity of the presentation. The DDME and backward DDME solvers additionally depend on $\{\hat{\eta}^{(i)}_{n_1,n_2,[k]}\}_{k=j'+1}^{j}$, which will be precomputed using $\{\{f_{m,(k)}^{(i)}\}_{m=1}^M\}_{k=0}^{j}$ and stored in the storage array $A$, hence the notation. Similarly, we suppress the inputs to the functions $\textsc{$D^{(i)}$\_Solver}$, $\textsc{$\hat{\eta}^{(i)}$\_Solver}$ and $\textsc{$\hat{\tilde{H}}^{(i)}$\_Solver}$ defined in the pseudocode whenever they are called, and their inputs are to be understood as corresponding to the inputs in the function definition unless specified otherwise. All sets of inputs in the function definitions are to be understood as running over all indices (e.g. $\{\hat{H}_m\}$ means $\{\hat{H}_m\}_{m=1}^M$). The time integral in \eqref{def. eta} and thus \eqref{eq. Krotov update rule 3} are approximated by Riemann sums.

\bibliography{references}

\end{document}